\documentclass[aps,pra,reprint,superscriptaddress, amsmath,amssymb,floatfix]{revtex4-1}
\usepackage{siunitx} 
\usepackage{subfigure} 
\usepackage[pdftex]{graphicx,epsfig}
\usepackage{graphics}
\usepackage{xcolor}
\usepackage{hyperref}
\newcommand\SRI{Sussex--RAL--ILL}

\newcommand{\Hz}{\mbox{Hz}}

\newcommand{\kV}{\ensuremath{\mathrm{kV}}}

\newcommand{\ecm}{\ensuremath{e\cdot\cm}}
\newcommand{\dn}{\ensuremath{d_{\textrm{n}}}}

\newcommand{\nT}{\mbox{nT}}
\newcommand{\pT}{\mbox{pT}}
\newcommand{\fT}{\mbox{fT}}

\newcommand{\cm}{\ensuremath{\mathrm{cm}}}
\newcommand{\mm}[1]{\ensuremath{\mathrm{mm}}}
\newcommand{\nm}{\ensuremath{\mathrm{nm}}}

\newcommand{\mum}[1]{\ensuremath{\micro \mathrm{m}}}

\newcommand{\magHg}{\ensuremath{{}^{199}\mathrm{Hg}}}
\newcommand{\magCs}{\ensuremath{{}^{133}\mathrm{Cs}}}

\renewcommand{\vec}[1]{\mathbf{#1}}

\newcommand{\Bonevec}{\vec{B}_{1}}

\newcommand{\Brfvec}{\vec{B}_\mathrm{RF}}
\newcommand{\nurf}{f_\mathrm{RF}}

\newcommand{\hl}[1] {{#1}}

\begin{document}

\title{Optically Pumped Cs{} Magnetometers Enabling a High-Sensitivity Search for the Neutron Electric Dipole Moment}

\author{C.~Abel}
\affiliation{Department of Physics and Astronomy, University of Sussex, BN1 9QH Brighton, England, United Kingdom}

\author{S.~Afach}
\affiliation{Paul Scherrer Institute, 5232 Villigen, Switzerland}
\affiliation{Eidgen\"ossische Technische Hochschule Z\"urich, Institute for Particle Physics and Astrophysics, 8093 Z\"urich, Switzerland}

\author{N.~J.~Ayres}
\affiliation{Department of Physics and Astronomy, University of Sussex, BN1 9QH Brighton, England, United Kingdom}
\affiliation{Eidgen\"ossische Technische Hochschule Z\"urich, Institute for Particle Physics and Astrophysics, 8093 Z\"urich, Switzerland}

\author{G.~Ban}
\affiliation{Normandie Universit\'e, ENSICAEN, UNICAEN, Centre National de la Recherche Scientifique IN2P3, LPC Caen, Caen, France}

\author{G.~Bison}
\email[corresponding author: ]{georg.bison@psi.ch}
\affiliation{Paul Scherrer Institute, 5232 Villigen, Switzerland}

\author{K.~Bodek}
\affiliation{Marian Smoluchowski Institute of Physics, Jagiellonian University, 30-348 Cracow, Poland}

\author{V.~Bondar}
\affiliation{Paul Scherrer Institute, 5232 Villigen, Switzerland}
\affiliation{Eidgen\"ossische Technische Hochschule Z\"urich, Institute for Particle Physics and Astrophysics, 8093 Z\"urich, Switzerland}
\affiliation{Instituut voor Kern- en Stralingsfysica, Katholieke Universiteit Leuven, 3001 Leuven, Belgium}

\author{E.~Chanel}
\affiliation{Laboratory for High Energy Physics, Albert Einstein Center for Fundamental Physics, University of Bern, 3012 Bern, Switzerland}

\author{P.-J.~Chiu}
\affiliation{Paul Scherrer Institute, 5232 Villigen, Switzerland}
\affiliation{Eidgen\"ossische Technische Hochschule Z\"urich, Institute for Particle Physics and Astrophysics, 8093 Z\"urich, Switzerland}

\author{C.~B.~Crawford}
\affiliation{University of Kentucky, Lexington, Kentucky 40506, USA}

\author{Z.~Chowdhuri}
\affiliation{Paul Scherrer Institute, 5232 Villigen, Switzerland}

\author{M.~Daum}
\affiliation{Paul Scherrer Institute, 5232 Villigen, Switzerland}

\author{S.~Emmenegger}
\affiliation{Eidgen\"ossische Technische Hochschule Z\"urich, Institute for Particle Physics and Astrophysics, 8093 Z\"urich, Switzerland}

\author{L.~Ferraris-Bouchez}
\affiliation{Universit\'e Grenoble Alpes, Centre National de la Recherche Scientifique, Grenoble INP, LPSC-IN2P3, Grenoble, France}

\author{M.~Fertl}
\affiliation{Paul Scherrer Institute, 5232 Villigen, Switzerland}
\affiliation{Eidgen\"ossische Technische Hochschule Z\"urich, Institute for Particle Physics and Astrophysics, 8093 Z\"urich, Switzerland}
\affiliation{Institute of Physics, Johannes Gutenberg University Mainz, 55128 Mainz, Germany}

\author{B.~Franke}
\altaffiliation{present address: TRIUMF, Vancouver, Canada}
\affiliation{Paul Scherrer Institute, 5232 Villigen, Switzerland}
\affiliation{Eidgen\"ossische Technische Hochschule Z\"urich, Institute for Particle Physics and Astrophysics, 8093 Z\"urich, Switzerland}

\author{W.~C.~Griffith}
\affiliation{Department of Physics and Astronomy, University of Sussex, BN1 9QH Brighton, England, United Kingdom}

\author{Z.~D.~Gruji\'c}
\affiliation{Physics Department, University of Fribourg, 1700 Fribourg, Switzerland}
\affiliation{Institute of Physics Belgrade, University of Belgrade, 11080 Belgrade, Serbia} 

\author{L.~Hayen}
\affiliation{Instituut voor Kern- en Stralingsfysica, Katholieke Universiteit Leuven, 3001 Leuven, Belgium}

\author{V.~H\'{e}laine}
\affiliation{Paul Scherrer Institute, 5232 Villigen, Switzerland}
\affiliation{Normandie Universit\'e, ENSICAEN, UNICAEN, Centre National de la Recherche Scientifique IN2P3, LPC Caen, Caen, France}

\author{N.~Hild}
\affiliation{Paul Scherrer Institute, 5232 Villigen, Switzerland}
\affiliation{Eidgen\"ossische Technische Hochschule Z\"urich, Institute for Particle Physics and Astrophysics, 8093 Z\"urich, Switzerland}

\author{M.~Kasprzak}
\email[corresponding author: ]{malgorzata.kasprzak@psi.ch}
\affiliation{Paul Scherrer Institute, 5232 Villigen, Switzerland}
\affiliation{Instituut voor Kern- en Stralingsfysica, Katholieke Universiteit Leuven, 3001 Leuven, Belgium}
\affiliation{Physics Department, University of Fribourg, 1700 Fribourg, Switzerland}

\author{Y.~Kermaidic}
\altaffiliation{present address: Max Planck Institut f\"ur Kernphysik, Heidelberg, Germany}
\affiliation{Universit\'e Grenoble Alpes, Centre National de la Recherche Scientifique, Grenoble INP, LPSC-IN2P3, Grenoble, France}

\author{K.~Kirch}
\affiliation{Paul Scherrer Institute, 5232 Villigen, Switzerland}
\affiliation{Eidgen\"ossische Technische Hochschule Z\"urich, Institute for Particle Physics and Astrophysics, 8093 Z\"urich, Switzerland}

\author{P.~Knowles}
\altaffiliation{present address: LogrusData, Toronto, Ontario, Canada}
\affiliation{Physics Department, University of Fribourg, 1700 Fribourg, Switzerland}

\author{H.-C.~Koch}
\affiliation{Paul Scherrer Institute, 5232 Villigen, Switzerland}
\affiliation{Institute of Physics, Johannes Gutenberg University Mainz, 55128 Mainz, Germany}
\affiliation{Physics Department, University of Fribourg, 1700 Fribourg, Switzerland}

\author{S.~Komposch}
\affiliation{Paul Scherrer Institute, 5232 Villigen, Switzerland}
\affiliation{Eidgen\"ossische Technische Hochschule Z\"urich, Institute for Particle Physics and Astrophysics, 8093 Z\"urich, Switzerland}

\author{P.~A.~Koss}
\affiliation{Instituut voor Kern- en Stralingsfysica, Katholieke Universiteit Leuven, 3001 Leuven, Belgium}

\author{A.~Kozela}
\affiliation{Henryk Niewodniczanski Institute of Nuclear Physics, Polish Academy of Sciences, 31-342 Cracow, Poland}

\author{J.~Krempel}
\affiliation{Eidgen\"ossische Technische Hochschule Z\"urich, Institute for Particle Physics and Astrophysics, 8093 Z\"urich, Switzerland}

\author{B.~Lauss}
\affiliation{Paul Scherrer Institute, 5232 Villigen, Switzerland}

\author{T.~Lefort}
\affiliation{Normandie Universit\'e, ENSICAEN, UNICAEN, Centre National de la Recherche Scientifique IN2P3, LPC Caen, Caen, France}

\author{Y.~Lemi\`ere}
\affiliation{Normandie Universit\'e, ENSICAEN, UNICAEN, Centre National de la Recherche Scientifique IN2P3, LPC Caen, Caen, France}

\author{A.~Leredde}
\affiliation{Universit\'e Grenoble Alpes, Centre National de la Recherche Scientifique, Grenoble INP, LPSC-IN2P3, Grenoble, France}

\author{A.~Mtchedlishvili}
\affiliation{Paul Scherrer Institute, 5232 Villigen, Switzerland}

\author{P.~Mohanmurthy}
\affiliation{Paul Scherrer Institute, 5232 Villigen, Switzerland}
\affiliation{Eidgen\"ossische Technische Hochschule Z\"urich, Institute for Particle Physics and Astrophysics, 8093 Z\"urich, Switzerland}

\author{M.~Musgrave}
\altaffiliation{present address: Massachusetts Institute of Technology, Cambridge, MA, USA}
\affiliation{Department of Physics and Astronomy, University of Sussex, BN1 9QH Brighton, England, United Kingdom}

\author{O.~Naviliat-Cuncic}
\affiliation{Normandie Universit\'e, ENSICAEN, UNICAEN, Centre National de la Recherche Scientifique IN2P3, LPC Caen, Caen, France}

\author{D.~Pais}
\affiliation{Paul Scherrer Institute, 5232 Villigen, Switzerland}
\affiliation{Eidgen\"ossische Technische Hochschule Z\"urich, Institute for Particle Physics and Astrophysics, 8093 Z\"urich, Switzerland}

\author{A.~Pazgalev}
\affiliation{Ioffe Physical Technical Institute,  194021 Saint Petersburg, Russia}

\author{F.~M.~Piegsa}
\affiliation{Laboratory for High Energy Physics, Albert Einstein Center for Fundamental Physics, University of Bern, 3012 Bern, Switzerland}

\author{E.~Pierre}
\affiliation{Paul Scherrer Institute, 5232 Villigen, Switzerland}
\affiliation{Normandie Universit\'e, ENSICAEN, UNICAEN, Centre National de la Recherche Scientifique IN2P3, LPC Caen, Caen, France}

\author{G.~Pignol}
\affiliation{Universit\'e Grenoble Alpes, Centre National de la Recherche Scientifique, Grenoble INP, LPSC-IN2P3, Grenoble, France}


\author{P.~N.~Prashanth}
\affiliation{Paul Scherrer Institute, 5232 Villigen, Switzerland}
\affiliation{Instituut voor Kern- en Stralingsfysica, Katholieke Universiteit Leuven, 3001 Leuven, Belgium}

\author{G.~Qu\'em\'ener}
\affiliation{Normandie Universit\'e, ENSICAEN, UNICAEN, Centre National de la Recherche Scientifique IN2P3, LPC Caen, Caen, France}

\author{M.~Rawlik}
\affiliation{Eidgen\"ossische Technische Hochschule Z\"urich, Institute for Particle Physics and Astrophysics, 8093 Z\"urich, Switzerland}

\author{D.~Rebreyend}
\affiliation{Universit\'e Grenoble Alpes, Centre National de la Recherche Scientifique, Grenoble INP, LPSC-IN2P3, Grenoble, France}

\author{D.~Ries}
\affiliation{Institut f\"ur Kernchemie, Johannes Gutenberg University Mainz, 55128 Mainz, Germany}

\author{S.~Roccia}
\affiliation{CSNSM, Universit\'e Paris Sud, Centre National de la Recherche Scientifique IN2P3, Universit\'e Paris Saclay, Orsay, France}
\affiliation{Institut Laue-Langevin, 38042 Grenoble, France}

\author{D.~Rozpedzik}
\affiliation{Marian Smoluchowski Institute of Physics, Jagiellonian University, 30-348 Cracow, Poland}

\author{P.~Schmidt-Wellenburg}
\affiliation{Paul Scherrer Institute, 5232 Villigen, Switzerland}

\author{A.~Schnabel}
\affiliation{Physikalisch Technische Bundesanstalt, D-10587 Berlin, Germany}

\author{N.~Severijns}
\affiliation{Instituut voor Kern- en Stralingsfysica, Katholieke Universiteit Leuven, 3001 Leuven, Belgium}

\author{R.~Tavakoli Dinani}
\affiliation{Instituut voor Kern- en Stralingsfysica, Katholieke Universiteit Leuven, 3001 Leuven, Belgium}

\author{J.~Thorne}
\affiliation{Department of Physics and Astronomy, University of Sussex, BN1 9QH Brighton, England, United Kingdom}
\affiliation{Laboratory for High Energy Physics, Albert Einstein Center for Fundamental Physics, University of Bern, 3012 Bern, Switzerland}

\author{A.~Weis}
\affiliation{Physics Department, University of Fribourg, 1700 Fribourg, Switzerland}

\author{E.~Wursten}
\email[corresponding author: ]{elise.wursten@cern.ch}
\altaffiliation{present address: CERN, Geneva, Switzerland}
\affiliation{Instituut voor Kern- en Stralingsfysica, Katholieke Universiteit Leuven, 3001 Leuven, Belgium}

\author{G.~Wyszynski}
\affiliation{Eidgen\"ossische Technische Hochschule Z\"urich, Institute for Particle Physics and Astrophysics, 8093 Z\"urich, Switzerland}
\affiliation{Marian Smoluchowski Institute of Physics, Jagiellonian University, 30-348 Cracow, Poland}

\author{J.~Zejma}
\affiliation{Marian Smoluchowski Institute of Physics, Jagiellonian University, 30-348 Cracow, Poland}

\author{G.~Zsigmond}
\affiliation{Paul Scherrer Institute, 5232 Villigen, Switzerland}

\date{\today}

\newpage

\begin{abstract} 
An array of sixteen laser-pumped scalar Cs{} magnetometers was part of the neutron electric dipole moment (nEDM) experiment taking data at the Paul Scherrer Institute in 2015 and 2016.
It was deployed to measure the gradients of the experiment's magnetic field and to monitor their temporal evolution. 
The originality of the array lies in its compact design, in which a single near-infrared diode laser drives all magnetometers that are located in a high-vacuum chamber, with a selection of the sensors mounted on a high-voltage electrode.
We describe details of the Cs{} sensors' construction and modes of operation, emphasizing the accuracy and sensitivity of the magnetic field readout. 
We present two applications of the magnetometer array directly beneficial to the nEDM experiment: (i)~the implementation of a strategy to correct for the drift of the vertical magnetic field gradient and (ii)~a procedure to homogenize the magnetic field.
The first reduces the uncertainty of the new nEDM result.
The second enables transverse neutron spin relaxation times exceeding 1500\,s, improving the statistical sensitivity of the nEDM experiment by about 35\% and effectively increasing the rate of nEDM data taking by a factor of 1.8.
\end{abstract}
\maketitle

\section{Introduction}
\label{sec:Intro}
%
The experimental search for a permanent electric dipole moment of the neutron (nEDM) has been an important topic of fundamental research since the early 1950s \cite{Pur50, Smi57}.
\hl{
Since then, the experimental sensitivity has been improved by more than six orders of magnitude.
The largest leap in sensitivity was due to the development of sources of ultracold neutrons (UCN) \cite{UCNsource, Golubbook} permitting the storage of neutrons within a material ``bottle" for hundreds of seconds \cite{Pendlebury1984}. 
This, in turn, created the requirement to keep experimental conditions, especially the magnetic field, stable over similar time spans, which resulted in the development of magnetometers placed close to \cite{Smith90} or within \cite{Green98} the storage bottle.
}
The experimental method applied to search for an nEDM with ultracold neutrons is based on a precise determination of the neutron spin precession frequency in static homogeneous parallel/antiparallel magnetic and electric fields by the Ramsey technique of \mbox{(time-)}separated oscillatory fields \cite{Ram50}. 
The statistical and systematic uncertainties of this method are strongly dependent on the (non)uniformity of the magnetic field $\vec{B}$ in which the neutrons precess.


This article is the second episode in a trilogy of papers that comprehensively treat the uncertainties in nEDM searches that originate from the inhomogeneity of the magnetic field.
The first episode \cite{Abe19a} gives a general introduction to the subject, defines the way we characterize gradients, and derives the relevant criteria for nEDM experiments. 
\hl{
In the second episode, \hl{ this paper, }we discuss the general approach to measure and compensate magnetic field gradients using an array of magnetometers. 
We describe in detail the specific implementation of this approach used in the 2015 and 2016 data runs of the nEDM experiment at the Paul Scherrer Institute (PSI).
The general concept and aspects of the implementation are applicable to other experiments where magnetic field homogeneity is a concern.
}
The third part will present the offline characterization of the magnetic field uniformity in the apparatus with an automated field-mapping device.
%


The nEDM apparatus at PSI is an upgraded version of the \SRI{} apparatus \cite{Bak14} that is equipped with two high-sensitivity systems for monitoring magnetic field changes, namely a \magHg{} co-magnetometer~\cite{Green98, Bak14} and an array of sixteen laser-pumped   Cs{} magnetometers~\cite{Weis05,Gro06}.
The PSI-nEDM experiment \cite{psiprl} was the first that used simultaneously a co-magnetometer and an array of external magnetometers during data taking. 
The Hg co-magnetometer employs an ensemble of spin-polarized \magHg{} atoms which occupy the same storage volume as the UCN, and whose spin precession frequency is used to correct for drifts of the magnetic field in every Ramsey cycle. 
The array of Cs{} magnetometers located above and below the storage chamber measures the spatial distribution of the magnetic field, allowing for control of the field homogeneity and extraction of the gradients across the neutron storage chamber.
The focus of this article is the implementation and application of the Cs{} magnetometer array.
In Section \ref{sec:nEDMexperiment} we describe the principle of the PSI-nEDM measurement with emphasis on the required magnetic field sensitivity and resolution of the magnetic field gradient.
Section \ref{sec:TheCsmagnetometerarray} provides a technical description of the Cs{} magnetometer array, including the design of the Cs{} magnetometers, their modes of operation and their performance in terms of magnetic field sensitivity and accuracy.
Section \ref{sec:Applications} details the applications of the Cs{} magnetometer array in the nEDM experiment.
A description of how to extract magnetic field gradients from the array field measurements is provided in 
Section \ref{sec:GzExtraction} and Section \ref{sec:T2opt} presents the procedure used to optimize the magnetic field.

\section{The \lowercase{n}EDM experiment at PSI}
\label{sec:nEDMexperiment}
\begin{figure}
\includegraphics[width=\linewidth]{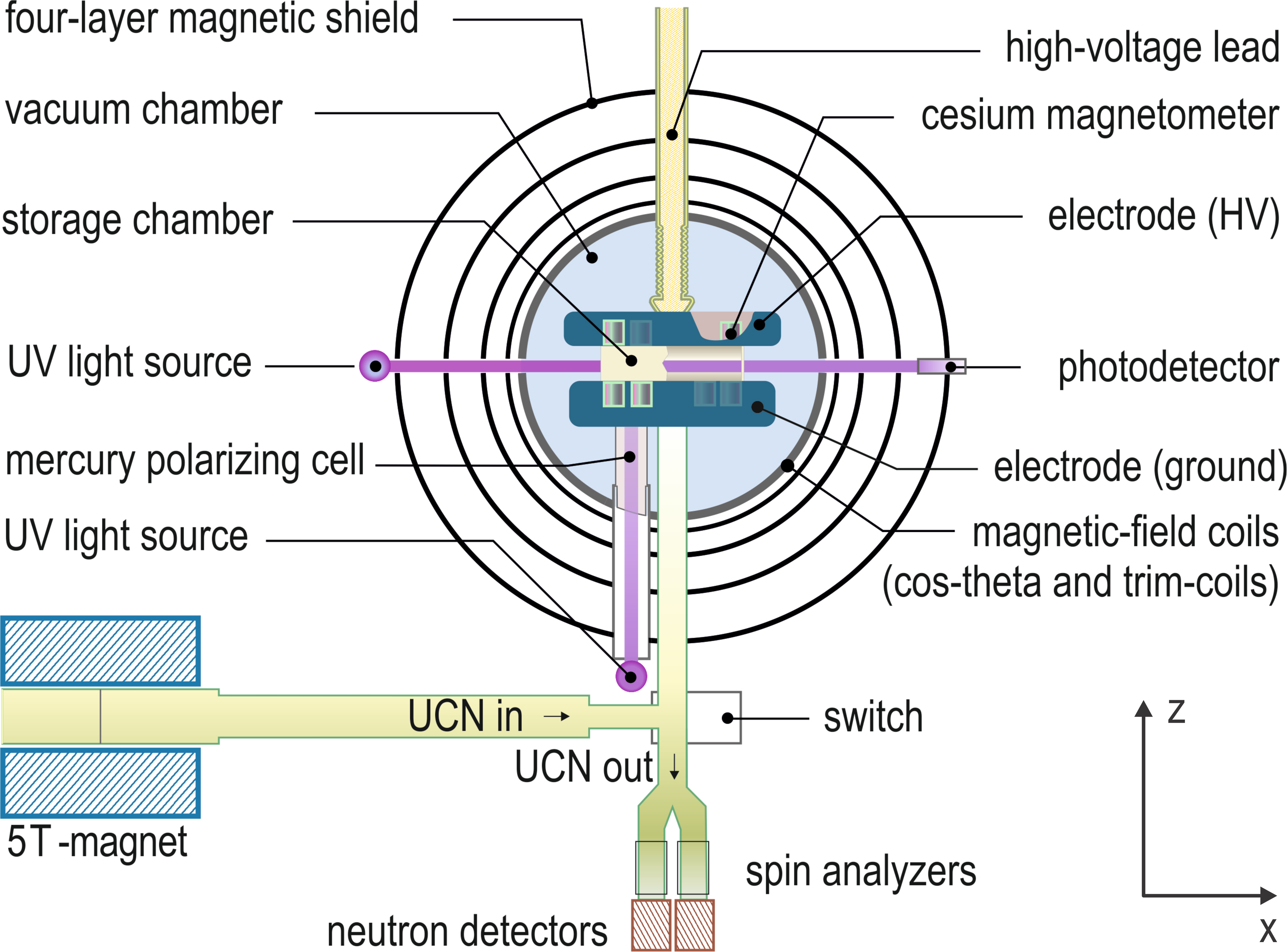}
\caption[Setup1]{Scheme of the nEDM apparatus. \hl{The magnetic and electric fields in the storage chamber are oriented vertically, each either parallel or anti-parallel to $z$.}}
\label{fig:setup_1}
\end{figure}

%
Figure~\ref{fig:setup_1} shows the general scheme of the PSI-nEDM experiment \cite{Bak11}, further called the `nEDM experiment'.
The cylindrical neutron storage chamber, \hl{ which also contains the \magHg{} co-magnetometer,}  consists of a polystyrene ring coated with deuterated polystyrene 
\cite{Bod} and aluminum end caps coated with diamond-like carbon \cite{Atc}.
The latter serve as high-voltage and ground electrodes, which can generate a vertical electric field of up to 15\,\kV/\cm{} in the chamber.
The height of the cylinder \hl{(i.e., the distance between the electrodes)} is 120\,mm, and the radius is 235\,mm.
The Cs{} magnetometers that measure the magnetic field gradients are mounted on the high-voltage and ground electrodes.
The storage chamber is located inside an aluminum vacuum chamber, onto which a cos-theta coil is wound.
\hl{The vacuum tank also supports a set of 30 trim-coils and the $B_1$ coils used to generate magnetic resonance pulses for the neutrons and the Hg atoms.}
The cos-theta coil produces a vertical, static magnetic field of $\approx \SI{1}{\micro T}$, while the set of trim-coils are used to homogenize the field and to apply specific field gradients when necessary.
The vacuum chamber is surrounded by a passive four-layer $\mu$-metal shield.
The whole setup is enclosed in an air-conditioned, temperature-stabilized wooden hut.
Three pairs of large ($\approx$ 8\,m$\times$6\,m) rectangular coils are mounted outside the hut and dynamically compensate the outer ambient field \cite{Afa14b}.
\hl{
The system attenuates fluctuations in the ambient field by factor of 5-50 in a bandwidth from DC to \SI{0.5}{Hz} which compensates the drop of passive shielding factor at small frequencies.
}

The operation of the apparatus during data taking with UCN was recently reviewed in \cite{Abe19b}. 
The effect of a finite nEDM  $d_{\textrm{n}}$ when the neutron is exposed to both an electric field $\vec{E}$ and a magnetic field $\vec{B}$ is an electric-field-dependent shift of the neutron spin precession frequency \hl{$f_n$}. 
Statistical uncertainties in the determination of that frequency by Ramsey's method \cite{Ram50} propagate to the sensitivity of the nEDM measurement
\begin{equation}
\sigma{}(d_{\textrm{n}})  =  \frac{\hbar }{2\,\alpha\, T \, E\, \sqrt{N} \sqrt{N_\mathrm{cycles}}},
\label{eq:nustatistical} 
\end{equation}
where $\alpha{}$ is the contrast of the Ramsey fringe, $T$ is the precession time, $E$ the electric field strength,  $N$ the number of detected neutrons in one Ramsey cycle, and $N_\mathrm{cycles}$ the number of such cycles.
\hl{In real measurements the statistical sensitivity is typically 10\% worse due to imperfections and data cuts.}
Details of this procedure are given in \cite{Harris07}.
The contrast $\alpha$ is determined by the transverse neutron spin depolarization time, and can be significantly improved by homogenizing the longitudinal (vertical) component of the magnetic field, as discussed in Section \ref{sec:T2opt}. 


The statistical sensitivity of the \SRI{} experiment \cite{Harris99,Bak06}, \hl{which led to the former best value for $\dn$ \cite{Pen15},}  was  $\sigma{}_{\textrm{day}}(d_\mathrm{n})\approx2\times10^{-25} \ecm$ per day ($N_\mathrm{cycles}=400$).
In the nEDM experiment at PSI \cite{psiprl} this value was improved by increasing $\alpha$ (see Section \ref{sec:T2optResults}), $E$, and neutron counting statistics and was on average $\sigma{}_{\textrm{day}}(d_\mathrm{n})\approx1.1\times10^{-25} \ecm$ per day ($N_\mathrm{cycles}=288$).
%

In order to keep the systematic uncertainty related to the control of the magnetic field and its gradients below the statistical sensitivity in Eq.~\eqref{eq:nustatistical}, the resolution of the magnetic field  measurement, \hl{$\sigma (B)$}, in one Ramsey cycle should be:
\begin{equation}
\sigma (B) \ll \frac{ E \sqrt{2 N_\mathrm{cycles}}~\sigma{}_{\textrm{day}}(d_{\mathrm{n}})}{\mu_n} ,
\label{eq:edmB}
\end{equation}
which gives $\sigma (B) \ll 0.5\,\pT$  for the PSI-nEDM experiment.
This resolution is provided by the \magHg{} co-magnetometer, whose spin precession frequency $f_{\textrm{Hg}}$ is used to monitor and correct for changes of the magnetic field from one Ramsey cycle to the next \cite{Bak14}. 
\hl{
Mercury, and specifically its isotope \magHg{}, was chosen because in its ground state it has no electronic contribution to the atomic spin. 
The atomic spin, which can be optically pumped and probed, is thus a pure nuclear spin with coherence times of up to hundreds of seconds. 
This permits to monitor the magnetic field during a Ramsey cycle with a coherent spin precession signal achieving a sensitivity that is on average better than 80~\fT{}.
Using the co-magnetometer signal as magnetic reference reduces the uncertainty of the neutron precession frequency due to magnetic field fluctuations to a few \% of the total uncertainty. 
}

%

\hl{All \magHg{} atoms are in the gas phase as the vapor pressure is much below the saturation pressure at room temperature. 
The atoms thus move with typical thermal velocities and sample the volume uniformly.
The ultracold neutrons, however, are noticeably affected by gravity because of their much lower velocity and thus preferentially inhabit the lower portion of the storage chamber
}
As a consequence, the ratio $\mathcal{R} ={f_{\textrm{n}}}/{f_{\textrm{Hg}}} $ is affected  by any vertical magnetic field gradient $\frac{\partial B_z}{\partial z}$ across the storage chamber. 
Adopting the notation of \cite{Abe19a},
\begin{equation}
\mathcal{R} =\frac{f_{\textrm{n}}}{f_{\textrm{Hg}}} = \frac{\gamma_{\textrm{n}}}{\gamma_{\textrm{Hg}}} \bigg(1 + \frac{G_{\textrm{grav}}\langle z \rangle}{B_0}+\delta_{\textrm{other}}\bigg) ,
\label{eq:Rratio}
\end{equation}
where $\gamma_{\textrm{n}}$ and $\gamma_{\textrm{Hg}}$ are the gyromagnetic ratios of the neutron and \magHg{} atom respectively, $G_{\textrm{grav}}$ is a combination of the relevant vertical gradients (see Section \ref{sec:GzExtraction}), $\langle z \rangle$ is the vertical displacement of the center of mass of the neutrons with respect to the center of the storage chamber, $B_0 = \langle B_z\rangle_{\textrm{Hg}}$ is the magnetic field averaged over the precession volume as measured by the \magHg{} co-magnetometer and $\delta_{\textrm{other}}$ encompasses all other effects that change the $\mathcal{R}$-ratio, such as, e.g., the motional false  EDM \cite{Afa15c} and the rotation of the Earth \cite{EarthRot}.
\hl{The positive $z$-direction is defined upwards with respect to gravity so that a negative value} is expected for the average displacement $\langle z \rangle$ of the neutrons.
The required resolution of the gradient measurements $\sigma{}(G_{\textrm{grav}})$ \hl{ for one Ramsey cycle} can be estimated in a similar way as $\sigma(B)$ leading to
\begin{equation}
\sigma{}(G_{\textrm{grav}}) \ll \frac{\sigma(B)}{|\langle z\rangle|}
\simeq\text{1.3\,pT/cm},
\label{eq:edmgz}
\end{equation}
using $\langle z\rangle = -0.38(3)$\,cm as determined in \cite{Abe19a}.

The temporal evolution of the magnetic field gradients was monitored with the array of sixteen Cs{} magnetometers installed close to  the precession chamber. 
\hl{This allowed corrections to be made for gradient drifts (Section \ref{sec:GzExtraction}) and the homogenization of the magnetic field using the variometer principle \cite{Ver06} (Sections \ref{sec:Variometer} and \ref{sec:T2opt}).}
The latter resulted in larger values for the contrast $\alpha$ leading to a 35\% increase in statistical sensitivity.

\section{The Cs{} magnetometer array}
\label{sec:TheCsmagnetometerarray}
%
This section describes the design, implementation and modes of operation of the Cs sensors installed above and below the precession chamber for monitoring magnetic field gradients.
\hl{The design decisions were guided by the requirement to minimize any potential interference between the Cs sensors and the neutron EDM measurement. 
We chose to operate the sensors at room temperature since temperature gradients can lead to electrical currents that disturb the magnetic field in the experiment.
Using Cs as the sensor medium combines two advantages in this situation: (i) Cs has the highest vapor pressure of all stable alkali metals and (ii) it has only one stable isotope, \magCs, with a large hyperfine splitting which suppresses interference from neighboring transitions. 
The sensors were operated in the $M_x$-mode \cite{Gro06, Ale, Weis16} which features a stable steady state due to the continuous magnetic resonance driven by an oscillating magnetic field. 
This weak field was suppressed by aluminum shielding cans and did not interfere with the neutron EDM measurement due to the large difference in resonance frequency (\SI{3.5}{kHz} for Cs vs. \SI{30}{Hz} for the neutrons).
}

\hl{
Similar sensor arrays have previously been used to measure the magnetic field generated by the human heart \cite{Array19,Array57}. 
For those biomagnetic measurements the performance of the array is limited by statistical uncertainties in the individual sensors. 
The sensors presented here are related to the ones used in \cite{Array19} but have been optimized for stability and accuracy since statistical uncertainties are not the limiting factor for the large integration times relevant in nEDM measurements.
}

\subsection{Design and implementation}
\label{sec:Design}
\begin{figure}
 \subfigure[]
 {\includegraphics[width=0.48\linewidth]{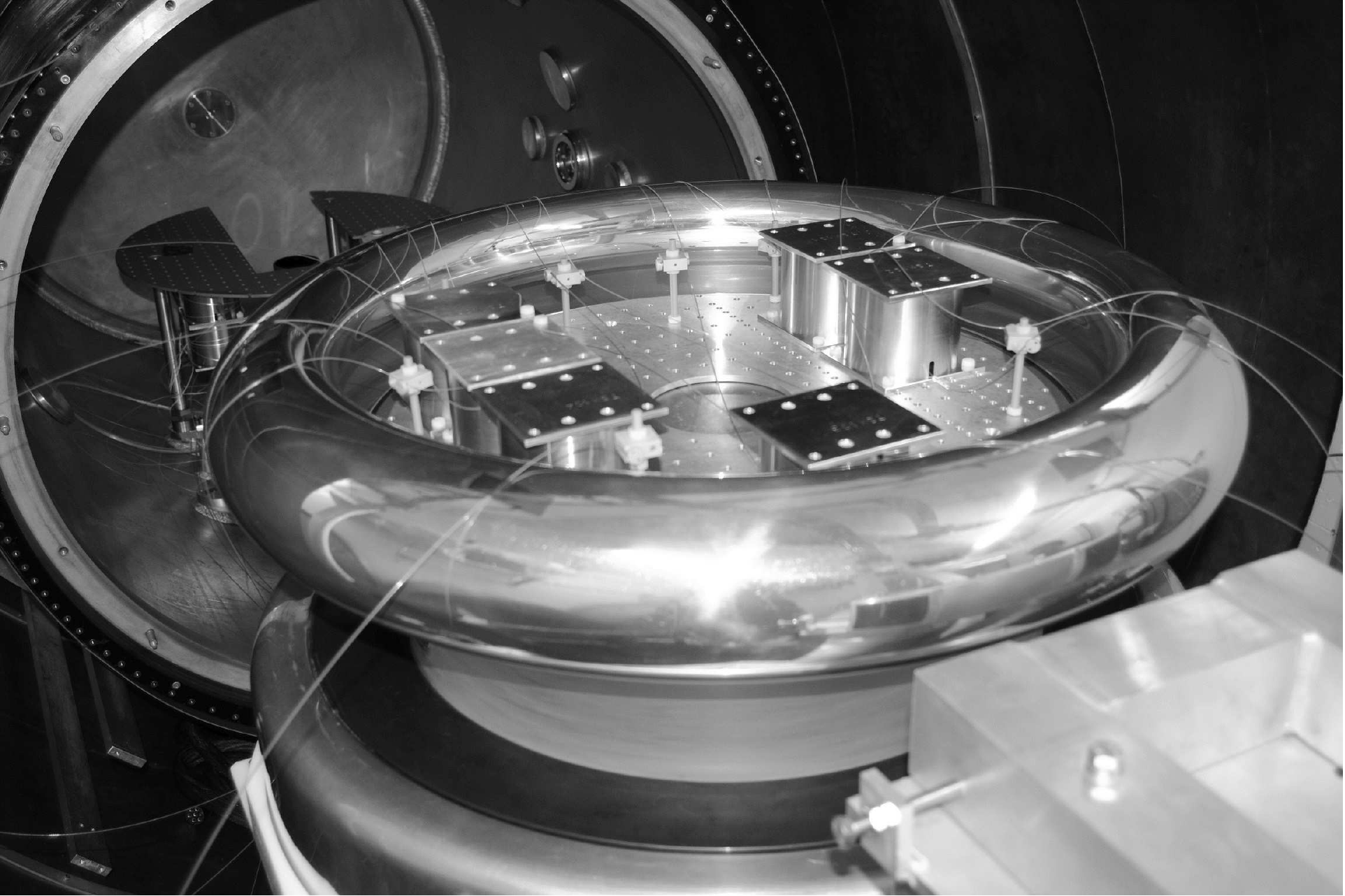}
 \label{fig:CsMaHV}}
 \hfill
 \subfigure[]
 {\includegraphics[width=0.48\linewidth]{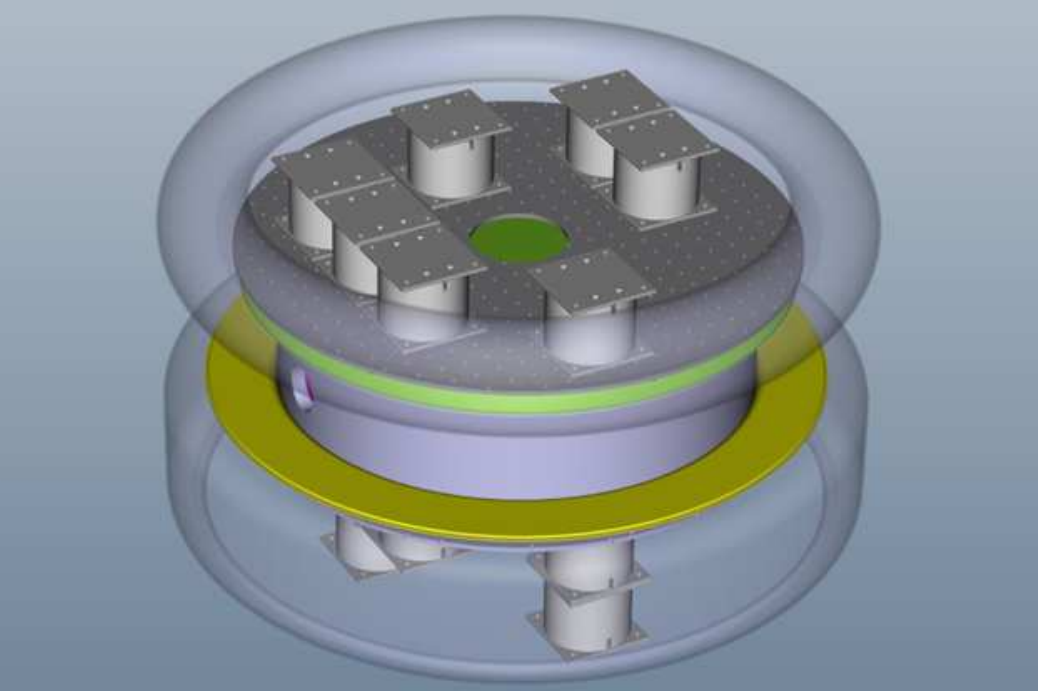}
 \label{fig:CsAll1}}
  \vfill
 \subfigure[]
 {\centering\includegraphics[width=0.38\linewidth]{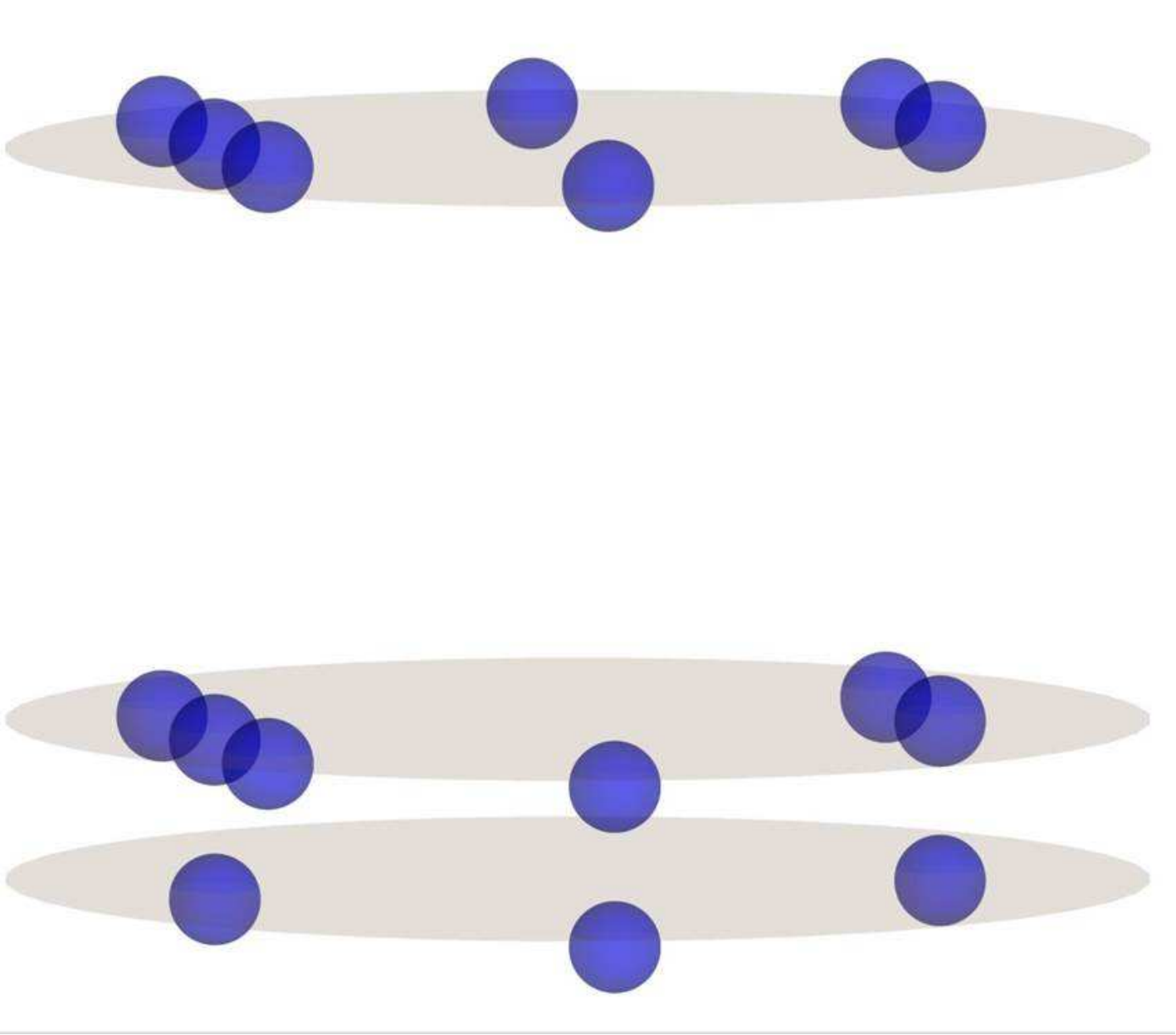}
 \label{fig:AllSensors}}
 \qquad{}
 \subfigure[]
 {\includegraphics[width=0.48\linewidth]{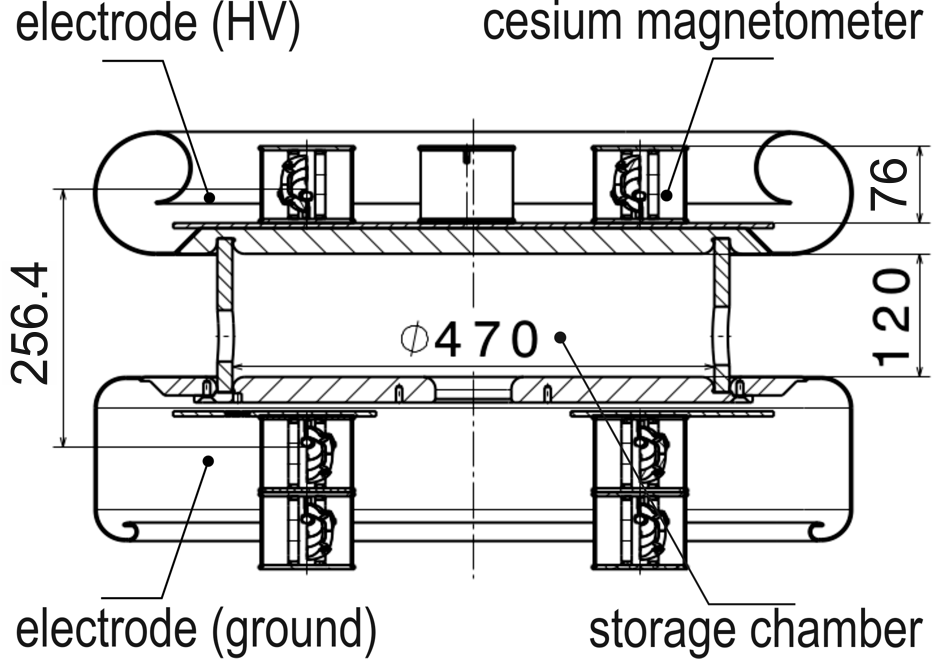}
 \label{fig:CsAll2}}
 \caption[Location of the Cs{} magnetometers in the nEDM experiment.]{Positions of the 16 Cs{}  magnetometers in the nEDM experiment. 
Each sensor is enclosed in an aluminum cylinder which suppresses the interaction of its RF-field with the neighboring Cs{} sensors. 
 (a)~Storage chamber removed from the vacuum chamber (in the background) with 6 HV-compatible Cs{} magnetometers installed on an aluminum plate fixed to the HV electrode with corona ring.
 (b)~Schematic view of the neutron storage chamber, the electrodes and  the Cs{} magnetometers. 
 (c)~The blue spheres indicate the positions of the Cs{} sensors, they are arranged in three layers above and below the storage chamber.
 (d)~Central vertical cut through (b) with  dimensions in mm.  
The vertical distance of the Cs{} sensors from the center of the  storage chamber is +127.9\,mm, -128.5\,mm, or -203.5\,mm, the 13 closest magnetometers thus being a factor of 2.8 closer to the center of the precession chamber in comparison to the $^{87}$Rb magnetometers in the earlier \SRI{} experiment 
\cite{Pen92}.
}
  \label{fig:CsMa}
\end{figure}

%
The magnetometer array consists of sixteen Cs{} sensors that are made of nonmagnetic materials and are vacuum-compatible.
The compact design allows their mounting close to the storage chamber.
The sixteen magnetometers are arranged in a three-layer gradiometer configuration with sensors located both above and below the storage chamber.
Seven sensors are installed on the high-voltage electrode, the centers of these sensors being 127.9\,\mm{} above the center plane of the neutron storage chamber.
Nine sensors are installed below the ground electrode.
They are arranged on two levels: 6 sensors are mounted on the aluminum plate directly below the ground electrode (128.5\,\mm{} below the center of the storage chamber),  while three more sensors are positioned in a plane located 75\,\mm{} lower, as shown in Fig.~\ref{fig:CsMa}. 
\hl{All sensors are placed with a position accuracy of about 0.5\,mm.}

\subsubsection{Principle of the Cs{} magnetometer}
\label{sec:Principleofthemagnetometer}
\begin{figure}
\includegraphics[width=\linewidth]{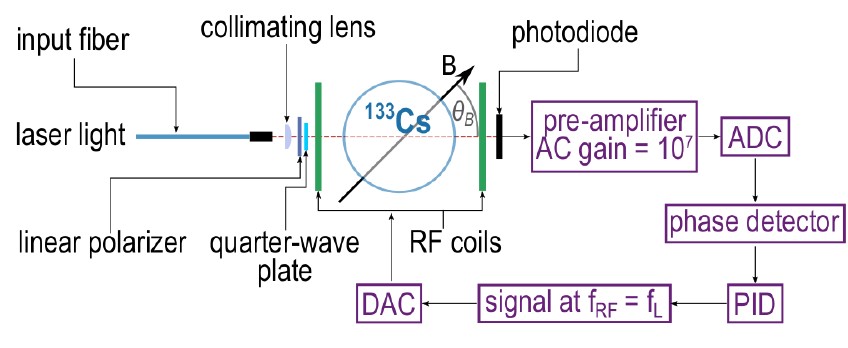}
\caption[sensor]{
Schematic of the  Cs{} magnetometers' main components and electronics as described in the text.}
\label{fig:sensor}
\end{figure}

%
The main components of a Cs{} magnetometer are shown in Fig.~\ref{fig:sensor}.
The actual field-sensing element of each sensor is an evacuated glass cell, with an inner diameter of $\sim$28~\mm{}, whose inner wall is coated by a thin layer of paraffin \cite{Cas}.
The Cs density in the cell is determined by the saturated vapor pressure of a metallic droplet of \magCs{} at room temperature.
The droplet is contained  in a sidearm connected to the main cell volume by a capillary.
The cesium atoms are spin-polarized by optical pumping using circularly polarized laser light whose frequency is resonant with the $F_g{=}4 \rightarrow F_e{=}3$ hyperfine component of the D$_1$ transition. 
The laser beam traverses the cell at an angle of 45$^\circ$ with respect to the magnetic field $\vec{B}$. 
The light from a frequency-stabilized laser is delivered to the sensor by a \SI{400}{\micro m} multimode fiber. 
Before entering the cell, the light is collimated by a lens and circularly polarized by a linear polarizer and a quarter-wave plate (Fig.~\ref{fig:sensor}).
The laser beam serves both to polarize the Cs{} atoms and to read out the precessing atomic spin polarization (optically detected magnetic resonance). 
When exposed to the magnetic field $\vec{B}$, the magnetic moment associated with the spin polarization  precesses at the Larmor frequency 
\begin{equation}
f_\mathrm{L} = \frac{\gamma_4}{2\pi} \|\vec{B}\|, \label{eq:Larmor}
\end{equation}
where $\gamma_4{} \simeq 2\pi \times 3.50$~\Hz/\nT~ is the gyromagnetic ratio of the $F{=}4$ hyperfine level of the cesium ground state. 
\hl{The spin precession can be either continuously driven by an oscillating magnetic field $\vec{B}_1$ or initiated by a magnetic resonance ($\vec{B}_1$) pulse (see Sec.~\ref{sec:PLLaccuracy}).
In both cases the $\vec{B}_1$ field is generated by a  Helmholtz-like pair of coils surrounding the Cs cell.
The coils were optimized to provide a homogeneous magnetic field over the volume of the Cs cell and  are historically named RF-coils, a convention we adopt here despite the low oscillation frequency of \SI{3.5}{kHz}.}
The precession of Cs{} atoms imposes an oscillation on the transmission of the laser light, which is detected on the photodiode.

All 16 magnetometers were operated  with light delivered by a single high-stability diode laser (Toptica, DL\,pro\,100) that was mounted in a dedicated housing in the temperature-stabilized room of the nEDM experiment. 
The laser frequency was actively locked to the $F_g{=}4\rightarrow F_e{=}3$ hyperfine component of the Cs{} D$_1$  ($6\textrm{S}_{1/2} \rightarrow 6\textrm{P}_{1/2}$) transition at $\sim$895~\nm{} using Doppler-free saturation absorption spectroscopy (Toptica, CoSy), which allowed us to keep the laser continuously in frequency lock for weeks.

The beam from this laser was divided into multiple beams by a splitter system which was directly attached to the main vacuum chamber of the nEDM apparatus.
The original beam was carried by a single \SI{400}{\micro m} multimode fiber to a beam homogenizer (SUSS MicroOptics) producing a flat-topped intensity profile of quadratic cross section.
The homogenized beam was then imaged onto a bundle of 36 fibers with \SI{400}{\micro m} core diameter whose flat-polished input ends are arranged into a square brass-epoxy holder with an aperture of 3$\times$3\,\mm{}$^2$.
Five of these fibers, including the four located at the corners of the bundle were used for monitoring purposes outside of the vacuum chamber.
The remaining 31 fibers ($\sim$4.5\,m long) were brought into the vacuum chamber, each with its own individual vacuum feedthrough.
\hl{In order to achieve stable transmission efficiencies, the fibers ran uninterrupted through  modified Swagelok feedthroughs which provided the vacuum sealing.}
Each fiber was terminated by a ferrule made of carbon-reinforced plastic that was inserted into the machined receptacle in the Cs{} sensor. 
On average, each output fiber carried  $\sim$1.4$\%$ of the input fiber's power.
%

\subsubsection{HV-compatible sensor modules}
\label{sec:HVcompatiblesensormodules}
%
%
\hl{The magnetometers mounted on the HV-electrode had to be fully opto-coupled.}
The  light transmitted by the cell was not detected by a photodiode mounted next to the cell, but rather coupled into  a \SI{3}{m} long  \SI{800}{\micro m} diameter multimode fiber carrying the light to a photodiode mounted on the grounded vacuum tank.
Tefzel$^\circledR$  (dielectric constant 2.6) was selected as a fiber coating in order to allow good electrical isolation of the sensor. 
\hl{
The RF signal driving the magnetic resonance was transmitted to the sensor by light generated by an IR LED (Lite-On Technology, model HSDL 4230) coupled to  a \SI{5}{m} long \SI{800}{\micro m} multimode fiber.
The plastic of the LED's casing was partly removed (down to a distance of $\sim$1--2\,\mm{} from the semiconductor die) and polished to optimize coupling into the fiber.
The light power had a constant and a sinusoidally modulated component which were converted to a current using a Si photodiode (Hamamatsu, model S6775-01) mounted near the sensor.
The photo current was sufficient to drive the RF-coils after it passed through a non-magnetic 470\,nF capacitor (WIMA 0.47 63/40) to suppress the DC component. 
}
\hl{
All sensors were operated with RF-field amplitudes approximately equal to the linewidth converted to magnetic field units,  $< \SI{4}{nT}$. 
}

\begin{figure}
\centering
\includegraphics[width=0.9\linewidth]{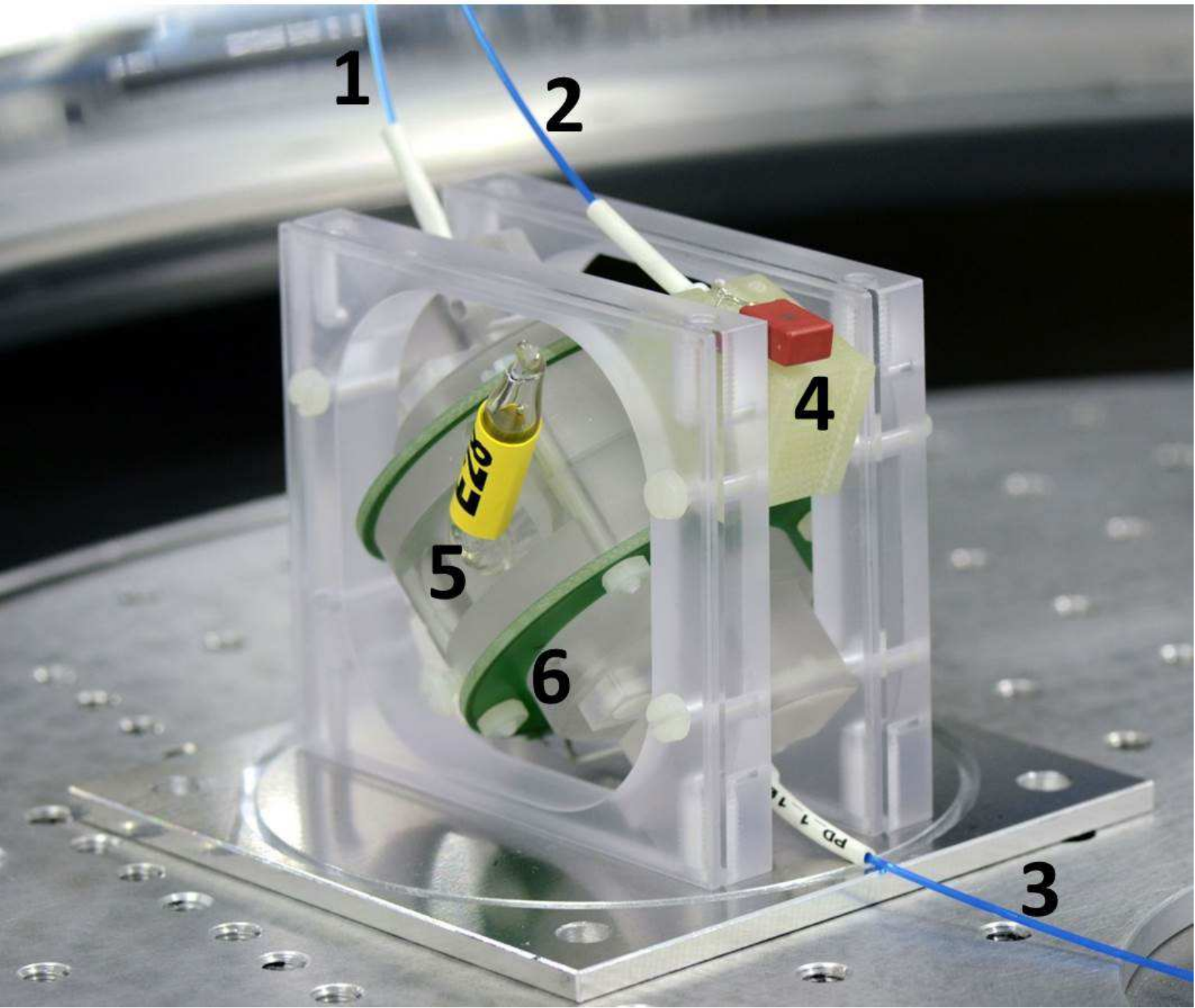}
\caption[CsHV]{HV-compatible magnetometer. 
The three fibers connected to the sensor provide the laser light (1), the RF signal (2), and collect the transmitted light (3). 
The Cs{} cell (of which only  the sidearm, 5, is visible) is placed in a poly-carbonate housing and surrounded by the RF coils printed on the (green) PCB boards (6). 
The photodiode and the capacitor forming the opto-coupler that drives the RF coils are mounted on a plastic holder (4).}
\label{fig:CsHV}
\end{figure}

\subsection{Phase-feedback mode of operation} 
\label{sec:pfmode}
%
\subsubsection{Description}
\label{sec:PLLdescription}
%
The magnetometer is operated in the $M_x$ configuration \cite{Gro06, Ale, Weis16} in which the precession of the Cs{} atoms' magnetization around $\vec{B}$ is continuously driven by a weak oscillating magnetic field $\Brfvec(t)=\Bonevec\sin(2\pi f_{\textrm{RF}} t)$.
\hl{
The $\Bonevec$ field is parallel to the wave vector of the laser beam,  $\Bonevec \parallel \vec{k}$, in order to avoid heading errors.
In this geometry, the shape and center of the magnetic resonance do not depend on the orientation of $\vec{B}$ with respect to $\vec{k}$ \cite{Weis16, Col}.
}

The light absorption by the Cs vapor depends on the projection of the atoms' magnetization onto $\vec{k}$.
\hl{The continuous magnetic resonance leads to a steady state magnetization which precesses at the driving frequency $f_{\textrm{RF}}$ and thus the transmitted light power has a component $\delta P(t)$ modulated at that frequency}
\begin{equation}
\delta P(t) = P_\mathrm{R}\sin(2\pi f_\mathrm{RF} \, t+\phi).
\end{equation}
Here $P_\mathrm{R}$ is the  modulation amplitude which depends on the light power, the degree of polarization, and the atomic absorption cross section.
The phase $\phi$ is the phase difference with respect to the driving field $\Brfvec$.
It has a characteristic resonant behavior \cite{Weis16}
\begin{equation}
\phi_E =\phi - \phi_0 =  - \arctan\left(\frac{f_\mathrm{RF} - f_\mathrm{L}}{\Gamma/2\pi}\right).
\label{eq:idealphase}
\end{equation}
\hl{Here $\Gamma = 1/T_1 = 1/T_2$ is the Cs spin relaxation rate, which is assumed to be isotropic.}
In absence of any additional phase shifts in the electronic circuits, the reference phase $\phi_0$ has the values of $\pm \pi/2$ depending on the direction of the magnetic field to be measured. 
The representation of the phase in Eq.~\eqref{eq:idealphase} is chosen such that the variable $\phi_E$ has a zero-crossing in the center of the resonance at $f_\mathrm{RF} = f_\mathrm{L} = \gamma_4 B/2\pi$. 
Close to that point $\phi_E$ is proportional to the difference between the driving frequency and the Larmor frequency. 
Its slope with respect to a change of the magnetic field magnitude can thus be expressed as
\begin{align}
\left.\frac{d \phi_E} {d B}\right|_{f_\mathrm{RF} = f_\mathrm{L}} 
&=\left.- \frac{d} {d B}  \arctan\left(\frac{f_\mathrm{RF} - \gamma_4 B/2\pi }{\Gamma/2\pi}\right) \right|_{f_\mathrm{RF} = f_\mathrm{L}} \nonumber\\
&= \frac{\gamma_4}{\Gamma}.
\label{eq:phaseslope}
\end{align}

The phase $\phi_E$ is determined by a digital signal processing (DSP) system that generates the driving frequency $f_\mathrm{RF}$ via a digital-to-analog converter and samples the photocurrent of the photodiode via an analog-to-digital converter\@. 
For this, the photocurrent which is proportional to the light power transmitted through the Cs cell is converted to a voltage by a transimpedance amplifier, prior to digitization.
The sampled voltage signal is then demodulated by a two-phase lock-in algorithm \cite{Weis16} that determines the amplitude of the oscillation and its phase.
The reference phase $\phi_0$ can be programmed via the digital interface of the DSP system which is also used to periodically read out the determined amplitude and phase values.

\begin{figure}
\centering
\includegraphics[width=0.9\linewidth]{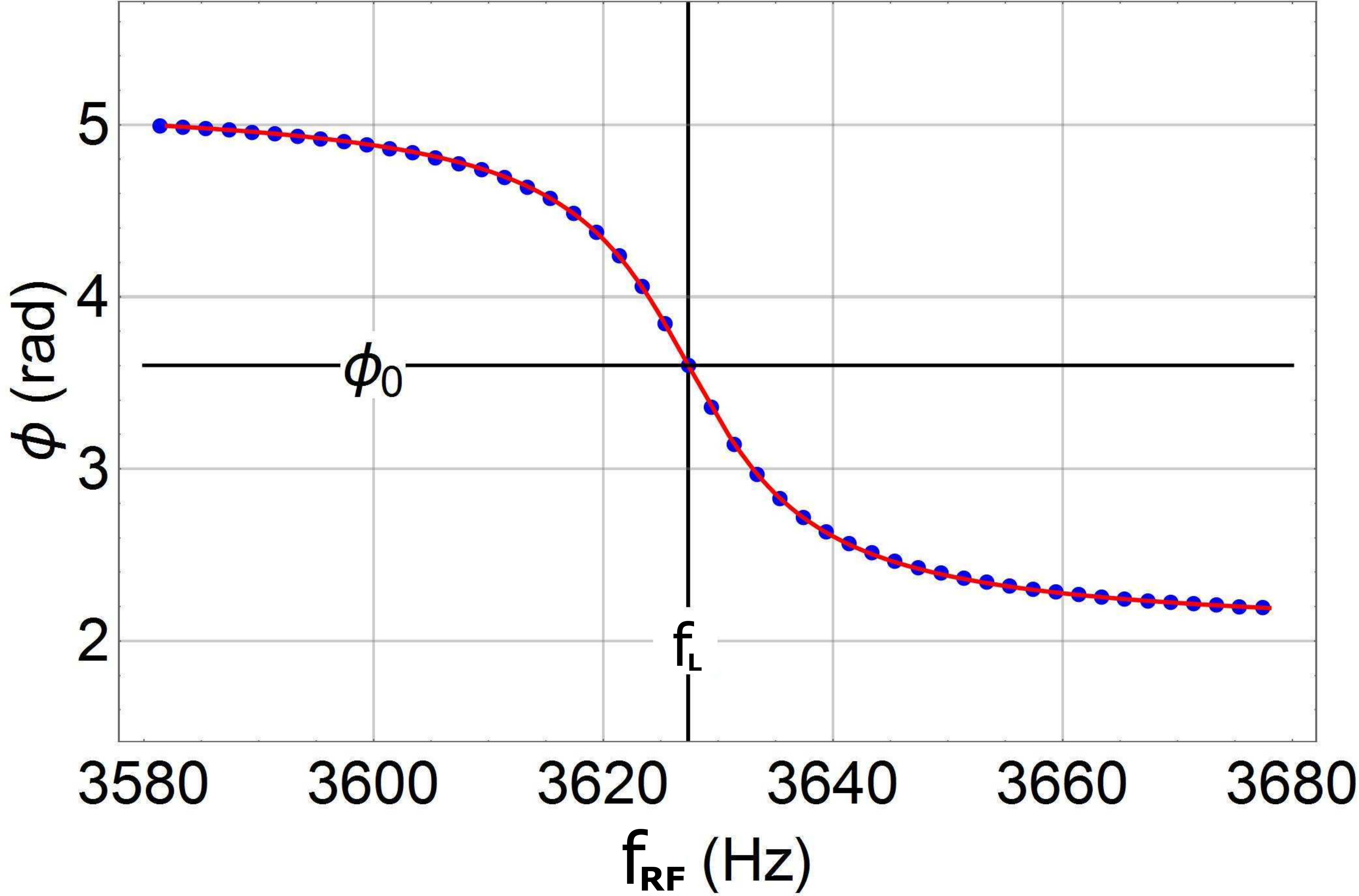}
\caption[The phase calibration curve.]{Typical calibration curve of a Cs{} sensor shown with the fit using Eq.~\eqref{eq:idealphase}.
The resulting fit parameters are: $\phi_0{=}$3.6032(8)~rad, $f_\mathrm{L}{=}$3619.980(8)~\Hz, and $\Gamma/2\pi{=}$5.358(7)~\Hz.}
\label{fig:phase}
\end{figure}

Figure \ref{fig:phase} shows a measurement of $\phi$ as a function of $f_\mathrm{RF}$. 
Such scans are used to determine the reference phase $\phi_0$ which is necessary to compute the shifted phase $\phi_E$. 
Phase shifts in the electronic circuits that are used in the generation of $f_\mathrm{RF}$ and the sampling of the photocurrent can cause changes in the reference phase $\phi_0$. 
The distinctive $\arctan$ line shape shown in Fig.~\ref{fig:phase} permits the determination of $\phi_0$ independently of external references. 
This procedure thus constitutes an internal calibration and is performed periodically.

In normal operation $f_\mathrm{RF}$ is not scanned. 
It is rather controlled by a servo algorithm that uses $\phi_E$ as its error signal.
If $\phi_0$ was correctly determined, keeping $\phi_E=0$ is equivalent to ensuring that  $f_\mathrm{RF} = f_\mathrm{L}$.
As a consequence, $f_\mathrm{RF}$, \hl{which is digitally synthesized in the DSP system,} becomes a measure for the magnetic field which is periodically sampled directly in the DSP system. 
This mode of operation using a feedback loop is similar to standard phase-locked-loop schemes. 
Here, however, a frequency offset does not result in a linearly changing error signal. 
Thus, in contrast to standard phase-locked loop systems, the error signal $\phi_E$ must not only be kept constant but also equal to zero in order to match $f_\mathrm{RF}$ and $f_\mathrm{L}$. 
This means that an offset $\Delta \phi_E$ in the determination of $\phi_E$ translates to an offset in the measured magnetic field according to Eq.~\eqref{eq:phaseslope}
\begin{equation}
\Delta B = \left(\frac{d \phi_E} {d B}\right)^{-1} \Delta \phi_E = \frac{\Gamma}{\gamma_4}\, \Delta \phi_E .
\label{eq:Berror}
\end{equation}

\subsubsection{Magnetometric sensitivity}
\label{sec:PLLsensitivity}

\begin{figure}
\centering
\includegraphics[width=\linewidth]{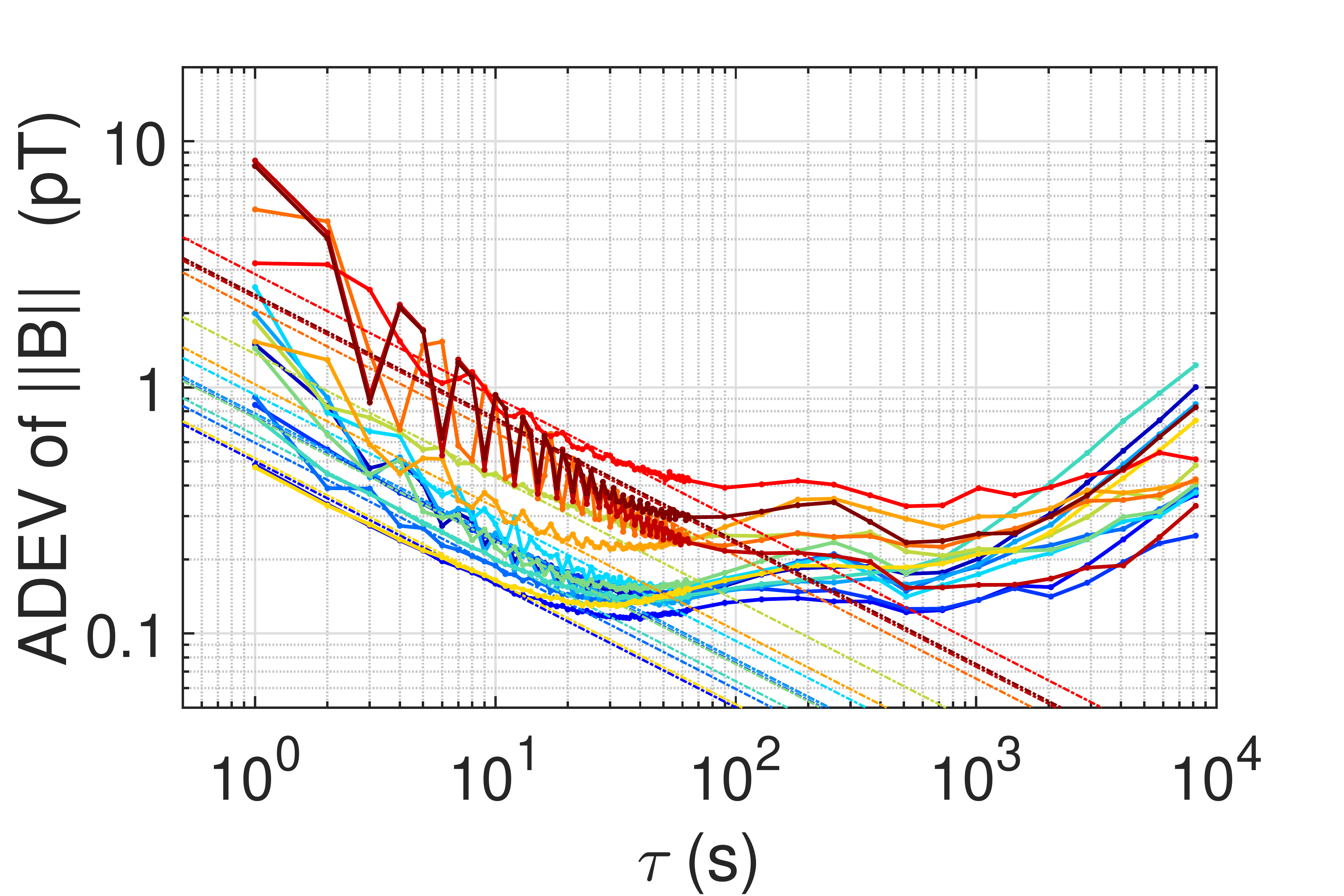}
\caption{Allan deviations of the magnetic field magnitude measured by 15 of the Cs{} magnetometers. The straight lines indicate the $\tau^{-1/2}$ behavior of pure white noise.  The oscillations that are visible for some sensors are caused by the RF field of a neighboring sensor, as explained in the text. }
\label{fig:ASDBz}
\end{figure}

The statistical uncertainty of the magnetic field measurement can be computed according to the propagation of noise from the sampled photocurrent $I^\mathrm{PD}$.
The phase noise spectral density is given by 
\begin{equation}
\rho{}(\phi) = \frac{\rho{}(I^\mathrm{PD})}{I^\mathrm{PD}_\mathrm{RF}} ,
\label{eq:phasesigma}
\end{equation}
where $I^\mathrm{PD}_\mathrm{RF}$ is the amplitude of the  oscillation in the photocurrent at the applied RF frequency.
Using Eq.~\eqref{eq:Berror} we find
\begin{equation}
\rho(B) =
\frac{ \Gamma \rho(\phi)}{\gamma_4}
=\frac{\Gamma }{\gamma_4}\frac{\rho(I^\mathrm{PD})}{I^{\mathrm{PD}}_\mathrm{RF}} .
\label{eq:sens2}
\end{equation}
In the shot noise limit, $\rho(I^\mathrm{PD}) = \sqrt{2\,e\,I^{\mathrm{PD}}_\mathrm{DC}}$ with $I^{\mathrm{PD}}_\mathrm{DC}$ the DC component of the photocurrent, the magnetometric sensitivity for all sensors used was better than $\rho(B) = \SI{50}{fT/\sqrt{Hz}}$ \hl{after the light power and $B_\mathrm{RF}$ amplitude were individually optimized for each sensor.}
The shot noise limit was used as the figure of merit for this optimization since it can be computed independently of the external magnetic noise which depended significantly on the changing experimental environment. 
During nEDM measurements the typical statistical sensitivity of the Cs magnetometers was $\rho(B) = \SI{750}{fT/\sqrt{Hz}}$.
\hl{The increase in statistical noise was due to the Johnson noise generated by the aluminum shielding cans} (thickness 2 mm) that had to be installed around each sensor to suppress interference from the $B_\mathrm{RF}$ fields of neighboring sensors. 
Even with the cans installed, a small amount of beating was observed due to the remaining interference. 
This is the reason why some magnetometers show a pronounced structure in the Allan deviations shown in Fig.~\ref{fig:ASDBz}. 
The resulting average sensitivity (including the beating effect) ranges from 0.75 to 8\,pT/$\sqrt{\mathrm{Hz}}$.

\subsubsection{Accuracy}
\label{sec:PLLaccuracy}
%
One can distinguish two types of effects that influence the accuracy of the Cs{} magnetometer. 
The first relates to inaccuracies in determining the Larmor precession frequency $f_{\mathrm{L}}$.
The second category includes all effects that change $f_\mathrm{L}$ itself, modifying the relation between the Larmor precession frequency and the magnetic field as given by Eq.~\eqref{eq:Larmor}.

Below follows a short discussion of both types, concluding with recommendations on how to keep the offsets as stable as possible, allowing for high relative accuracy of the magnetic field reading.

\begin{figure}
\centering
 \subfigure[]
{\includegraphics[width=0.48\linewidth]{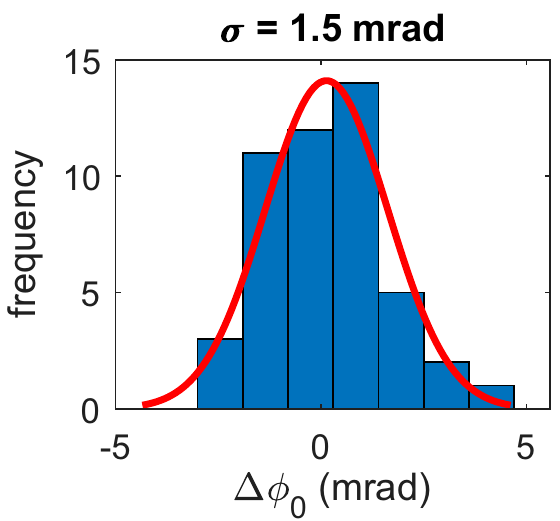}
\label{fig:phiDiff}}
 \hfill
 \subfigure[]
 {\includegraphics[width=0.48\linewidth]{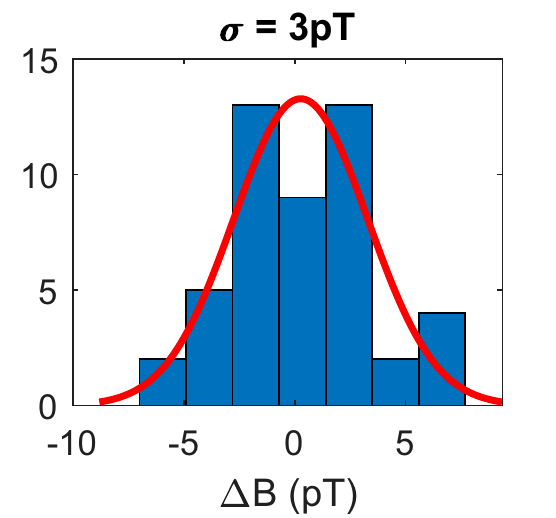}
 \label{fig:phaseBdiff}}
 \caption{(a)~Histogram of the difference in extracted reference phase $\phi_0$ between two consecutive calibrations.
 The typical time between the two calibrations is 1 to 4 days.
  (b)~The corresponding offset in the magnetic field reading, as calculated by Eq.~\eqref{eq:Berror}.
  The distributions in (a) and (b) are not identical, as the width $\Gamma$ in the conversion factor depends on the light intensity, which varies from sensor to sensor.
	\hl{Over days of data taking the values for $\Gamma$ were typically stable to better than 5\%. During the whole two year-data taking period, all sensors had values of $\Gamma/ 2\pi$ between \SI{4}{Hz} and \SI{17}Hz.}
   }
\label{fig:phasedrift}
\end{figure}

As the extraction of the Larmor precession frequency relies heavily on the reference phase $\phi_0$, any drift of $\phi_0$ without recalibration will worsen the accuracy of the sensor. 
Such drifts can occur due to temperature-related effects in the electronics or when, for example, the laser intensity changes \cite{JariThesis}  and thus the capacitance of the photodiode.  
In order to quantify such drifts in the nEDM Cs{} magnetometer array, we have performed calibrations before and after each nEDM run, typically 1 to 4 days apart.
Figure \ref{fig:phiDiff} shows a histogram of the extracted phase change $\Delta\phi_0 = \phi_{0,\textrm{after}} - \phi_{0,\textrm{before}}$ for one of the sixteen sensors.
The typical change of reference phase is on the order of 1 to \SI{2}{m rad}.
An uncorrected drift $\Delta\phi_0$ of the on-resonance phase during phase-feedback operation results in an offset in the magnetic field measurement according to Eq.~(\ref{eq:Berror}).
Figure \ref{fig:phaseBdiff} shows the results of converting the phase differences in Fig.~\ref{fig:phiDiff} to offsets in the magnetic field reading. 
The standard deviation of the magnetic field reading offset depends on the sensor properties and ranges from 1 to maximum 7\,pT.
This is of the same order of magnitude as the inherent uncertainty provided by the calibration procedure itself, which is about 1\,pT.
%

Regarding the second category of inaccuracies, there are several effects that modify the Larmor precession frequency, or to be more precise, the energy separation of adjacent Zeeman sublevels of the $F{=}4$ ground state of the \magCs{} atoms.
The resonance frequency that is measured by the Cs{} magnetometer in phase-feedback mode is a weighted average of the energy differences between the $m$ and $m+1$ magnetic sublevels.
In a system without laser interaction, the energy levels are the eigenvalues of the Cs{} ground state Hamiltonian containing the hyperfine interaction $A\,\vec{J}\cdot\vec{I}$ between the electronic spin $\vec{J}$ and the nuclear spin $\vec{I}$, and the interaction of the magnetic moment with the applied magnetic field $\vec{\mu}\cdot\vec{B}$.
Applying perturbation theory to first order in $\mu B/A$ (for $\mu B$ small compared to the scale given by the hyperfine structure constant $A$) then yields the linear Zeeman level splitting.
The exact solution for this $J=1/2$ system is given by the Breit-Rabi equation \cite{BreitRabi}.
For a magnetic field of \SI{1}{ \micro T}, the nonlinear terms in the Zeeman effect result in a maximum deviation equivalent to 3\,pT for neighboring magnetic sublevels, giving an upper limit on the inaccuracy due to nonlinear Breit-Rabi splitting.
%

 A second effect of this category has to do with the use of a nonrotating driving field $\vec{B}_{\textrm{RF}} = \vec{B}_1 \sin(2\pi \nurf t)$. 
The nonrotating field produces a Bloch-Siegert shift \cite{BlochSiegert,Sudyka}, which shifts the resonance by
\begin{equation}
\frac{\left(B_{1}\sin\theta_B\right)^2}{16B_0}= \frac{B_1^2}{32B_0}\approx 0.5\,\textrm{pT} ,
\label{eq:BlochSiegert}
\end{equation}
as the RF field of \SI{4}{nT} makes an angle of $\theta_B=\pi/4$ with the main $B_0$ field of \SI{1}{\micro T}.

Another interaction that modifies the energy of the magnetic sublevels is the AC Stark shift induced by the coherent laser light, otherwise known as the virtual light shift \cite{Happer}. 
It entails an interaction $\vec{d}\cdot\vec{E}$ between the electric dipole moment operator $\vec{d}$ of the Cs{} atoms and the oscillating electric field $\vec{E}$ of the laser light.
Apart from modifying the hyperfine splitting and the common energy of all levels, it also produces a linear splitting and a quadratic splitting of the magnetic sublevels.
The former is called a vector light shift, the latter a tensor light shift.
The vector light shift can be interpreted as an effective magnetic field that is oriented along the direction of the laser beam for $\sigma^+$ light.
As the laser light propagates at an angle of 45$^\circ$ with respect to $\vec{B}$, this effective magnetic field will add or subtract to the magnitude of the main magnetic field, depending on the direction of $\vec{B}$.
Both the vector and the tensor light shift in the $F_g{=}4$ ground state depend linearly on the intensity of the light and have a dispersive line shape relative to the laser detuning around each hyperfine transition. 
Although the dispersive function vanishes when the laser frequency is resonant with the respective transition $F_g{=}4 \rightarrow F_e{=}3$, the light shift itself does not, as the dispersive function of the neighboring transition $F_g{=}4 \rightarrow F_e{=}4$ is quite broad and nonzero at that laser frequency.
In order to determine the size of this effect in the nEDM experiment, dedicated measurements were done by changing the intensity of the light in a controlled way and scanning the detuning of the laser around the $F_g{=}4 \rightarrow  F_e{=}3$ transition.
To avoid the inaccuracy issues of the first type, the magnetometers were run in the free spin precession (FSP) mode \cite{Gru15,Afach15}.
\hl{They could be operated in FSP mode without changing the sensor hardware or the laser power. 
The waveform of the signal driving the RF-coils was changed to a burst which alternates between RF-pulses and periods of zero RF amplitude. 
During the periods without RF field the ensemble spin precesses freely while the constant laser interaction pumps it slowly to an equilibrium state parallel to $\vec{B}$.
The RF-pulses were tuned to flip the accumulated spin polarization by approximately 90$^\circ$ to the plane perpendicular to $\vec{B}$.
During the next free precession period of about \SI{50}{ms} the laser, which is oriented at 45$^\circ$ with respect to $\vec{B}$, 
probes the spin component parallel to $\vec{k}$, which contains both the precessing signal of the spin component perpendicular to $\vec{B}$ and the growing spin polarization created along $\vec{B}$ due to optical pumping.
}
The advantage of operating the magnetometer in the FSP mode is that one directly detects the Larmor spin precession frequency $f_{\textrm{L}}$ of the Cs{} atoms.
These FSP studies \cite{EliseThesis} have shown that the sensors display shifts ranging from $\pm10$\,pT to $\pm50$\,pT at their typical light intensities, which are correlated to the light intensity, depend on the laser detuning and indeed change sign as the magnetic field is reversed.
The FSP mode of operation was only used to test the Cs magnetometers since the pulse repetition frequency is close to the Larmor frequency of the \magHg{} atoms. 
Oscillating magnetic fields with frequency components close to the resonance frequency can cause changes in the Larmor precession of the \magHg{} atoms via the Ramsey-Bloch-Siegert shift which was not acceptable during nEDM data taking. 
Recent implementations of the FSP mode avoid interference via the RF field by using all-optical designs \cite{Gru15}.

A fourth effect that modifies the Hamiltonian of the atom-light system is due to spin-exchange collisions between the \magCs{} atoms \cite{Grossetete,SpinExchRelax}.
The frequency shift operator contains a term proportional to $\vec{S}\cdot \langle \vec{S}\rangle$, where $\vec{S}$ is the electron spin of the Cs{} atoms.
This effect scales with the number density of the alkali atoms \cite{SpinExchange} and is therefore exponentially dependent on temperature.
The exact implications for our magnetometer are not yet fully understood theoretically, but preliminary measurements comparing the precession frequency in different parts of the FSP signal (and thus at different directions of $\langle \vec{S}\rangle$) seem to indicate that the effect is smaller than 30\,pT for all sixteen sensors \cite{EliseThesis}.

An overview of the effects discussed above is given in Table~\ref{tab:Systematics}. Combining the values of the different effects, the absolute accuracy of the sensors adds up to be in a range from 45 to 90\,pT. 
For the purpose of measuring drifts of the vertical magnetic field gradient $G_{\textrm{grav}}$, the absolute accuracy of the magnetometers is not crucial, but it is important that the relative reading offsets of all sensors remain stable in time.
It is therefore recommended to keep the light intensity sufficiently stable to avoid drifts in the reference phase and to keep the light shift in check. 
Additionally, large changes in temperature should be avoided, both for the stability of the electronics and the spin exchange effect.
The achieved stability in the nEDM experiment was significantly better than the requirements for time scales up to \SI{10000}{s} as discussed in section \ref{sec:GzExtraction}.

\begin{table}
\caption{Overview of effects that relate to inaccuracies in determining the Larmor precession frequency $f_{\mathrm{L}}$ (line 1), or that change $f_\mathrm{L}$ itself (lines 2 to 5) thereby modifying the relation between the Larmor precession frequency and the magnetic field as given in Eq.~\eqref{eq:Larmor}.}
\centerline{
\begin{tabular}{l c}
\hline
\hline
Effect & size (pT) \\
\hline
Reference phase drifts & 1 to 7 \\
Quadratic Zeeman splitting & 3 \\
Bloch-Siegert shift & 0.5 \\
Vector light shift & 10 to 50 \\
Spin exchange & $<$30 \\
\hline
\hline
\end{tabular}
}
\label{tab:Systematics}
\end{table}

\subsection{Variometer method}
\label{sec:Variometer}
%
The array of Cs{} magnetometers can be used to obtain the vector components of the magnetic field by applying the variometer principle \cite{Ver06}.
The implementation of this method will be explained in Section \ref{sec:VarioWp}, its sensitivity and accuracy will be discussed in Sections \ref{sec:VarioSensitivity} and \ref{sec:VarioAccuracy} respectively.
%

\subsubsection{Working principle}
\label{sec:VarioWp}
%
The variometer method consists of applying a well known magnetic field $\vec{B}_{\textrm{T}}$ transverse to the main magnetic field of \SI{1}{\micro T}.
Using the Cs{} magnetometers in phase-feedback mode to measure the magnitude of the total magnetic field, the additional transverse magnetic field changes the magnitude to:
\begin{equation}
\|\vec{B}_{0}+ \vec{B}_{\textrm{T}} I \|^2 =\|\vec{B}_{0}\|^2+ 2 \vec{B}_{0}\cdot\vec{B}_{\textrm{T}} I+  \|\vec{B}_{\textrm{T}}\|^2 I^2,  
\label{eq:Parabola}
\end{equation}
where $\vec{B}_{0}$ represents the main magnetic field, $I$ the current applied to the transverse coil,  and $\vec{B}_{\textrm{T}}$ the field produced by this transverse coil at the position of the Cs{} magnetometer when applying one unit of current.
Probing the field magnitude with a set of different currents, one can extract $\|\vec{B}_{0}\|$, $\|\vec{B}_{\textrm{T}}\|$ and $\vec{B}_{0}\cdot\vec{B}_{\textrm{T}}$ from the quadratic behavior of $\|\vec{B}_{0}+\vec{B}_{\textrm{T}}I \|^2$ as a function of the current.
The scalar product $\vec{B}_{0}\cdot\vec{B}_{\textrm{T}}$ contains the angle between the applied transverse magnetic field and the main field $\vec{B}_0$.
Projecting on two known transverse magnetic field directions, one can reconstruct the direction of $\vec{B}_0$.

\begin{figure}
\centerline{\includegraphics[width=\linewidth]{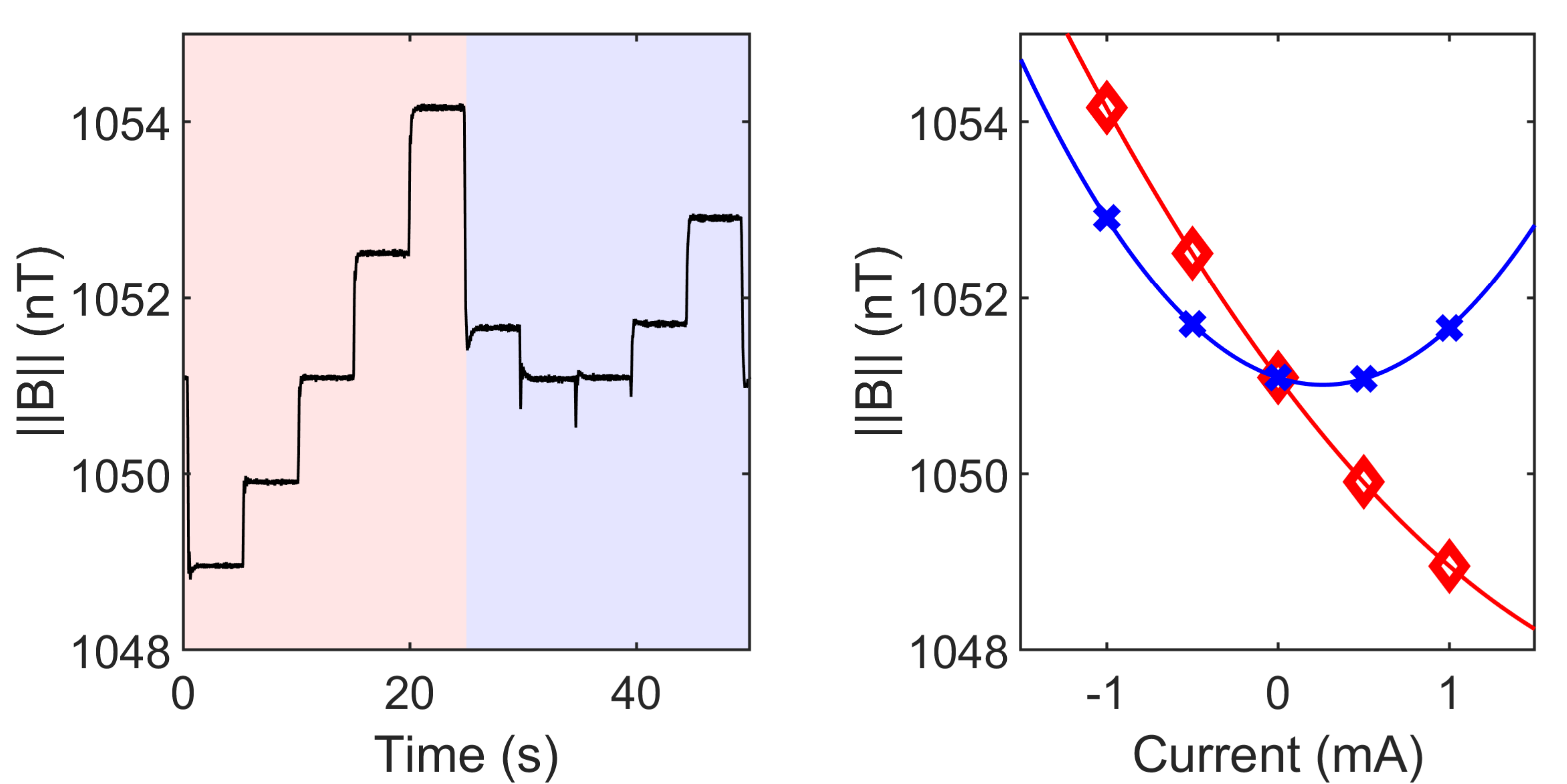}}
\caption{ 
On the left: the response of one of the Cs magnetometers to a current pattern of 1, 0.5, 0, -0.5 and -1\,mA in steps of \SI{5}{s}, first applied to a coil in the $x$-direction (0--25\,s indicated in red), then to a coil in the $y$-direction (25--50\,s indicated in blue). The main field of \SI{1.051}{\micro T} is maintained along the $z$-direction. A current of 1\,mA corresponds to an applied field of about \SI{50}{nT}. On the right: the corresponding parabolic behavior of the magnitude as a function of the applied current to a coil in the $x$-direction (red diamonds) and a coil in the $y$-direction (blue crosses).
}
\label{fig:CsMResponse}
\end{figure}

An example of the readout of a Cs{} sensor during the application of the variometer method is shown in Fig.~\ref{fig:CsMResponse}. 
Here, a sequence of five equally spaced currents is applied for five seconds each, first to a coil in the $x$-direction, then to a coil in the $y$-direction, whereas the main magnetic field is maintained in the $z$-direction.
The currents are applied with an Agilent 33500B function generator, using a resistor of 10\,k$\Omega$ in series with the transverse coils to convert the voltage generated by the function generator to a proportional current.
In order to avoid magnetization of the $\mu$-metal shield, the maximal current $I$ is chosen such that the transverse field is about a factor of 20 smaller than the main magnetic field of \SI{1}{\micro T}.
This results in a change of the magnetic field magnitude by typically 5\,nT.
As the Cs{} magnetometer is run in the phase-feedback mode, the reaction of the sensor to this sudden change of the magnetic field is not instantaneous, but has a time constant of a few 100\,ms, depending on the parameters of the stabilizing PID algorithm.
Consequently, the ramping parts of the signal have to be cut when averaging the magnitude over one current setting, effectively increasing the measurement uncertainty calculated in Section \ref{sec:VarioSensitivity}.
%

In order to extract the vector components of the main magnetic field, knowledge of the direction of the applied transverse field is crucial. 
The coils that are used to generate $\vec{B}_{\textrm{T}}$ are normally used for applying the UCN and \magHg{} $\pi/2$ spin-flip pulses in the nEDM experiment.
The magnetic fields produced by these coils were measured in 2014 with a nonmagnetic mapping device \hl{(the topic of the third episode in this trilogy)} consisting of a three-axis fluxgate magnetometer mounted on a trolley.
The trolley could move along a horizontal arm, which itself could rotate along a vertical axis and move up and down along the same vertical axis.
Scanning the volume in discrete steps, the magnetic field map can be reconstructed from the corresponding fluxgate readings \cite{GWmap}.
The resulting accuracy of these field maps at the specific Cs{} magnetometer positions is about 1\,nT on each magnetic field component for a 50\,nT total field produced by the coil. 
%

This 2\% inaccuracy of the field maps translates into a similar inaccuracy of all three vector components of $\vec{B}_0$ if the extraction is based purely on the two transverse projections. 
For this reason, we additionally include the fact that the magnetic field is predominantly homogeneous and assume that the $B_{0z}$ component of the main field is closely approximated by the field magnitude $B_{0z}=\pm\|\vec{B}_0\|$ (true at the tens-of-pT level), with the sign being determined by the set $B_0$ direction.
Using this approximation, one can extract $B_{0x}$ and $B_{0y}$ by solving the following set of equations:
\begin{equation}
\begin{bmatrix}
\vec{B}_{0}\cdot\vec{B}_1 - B_{0} B_{1z}  \\
\vec{B}_{0}\cdot\vec{B}_2 - B_{0} B_{2z}
\end{bmatrix}
=
\begin{bmatrix}
B_{1x} & B_{1y}\\
B_{2x} & B_{2y}
\end{bmatrix}
\begin{bmatrix}
B_{0x}\\
B_{0y}
\end{bmatrix}\label{eq:BtExtraction} ,
\end{equation}  
where $\vec{B}_1$ and $\vec{B}_2$ are the two applied transverse fields.
To take into account slight differences in applied currents during the maps and the variometer measurement, the $\vec{B}_1$ and $\vec{B}_2$ maps are scaled using the $\|\vec{B}_{\textrm{T}}\|^2$ parameter from the quadratic fit in Eq.~\eqref{eq:Parabola}.
Matrix inversion of Eq.~\eqref{eq:BtExtraction} yields
\begin{equation}
\begin{split}
B_{0x} &= \frac{B_{2y}\left(\vec{B}_{0}\cdot\vec{B}_1 - B_{0} B_{1z}\right) - B_{1y} \left(\vec{B}_{0}\cdot\vec{B}_2 - B_{0} B_{2z}\right) }{B_{1x} B_{2y} - B_{2x} B_{1y}}\\
&\approx  \frac{\vec{B}_{0}\cdot\vec{B}_1 - B_{0} B_{1z} }{B_{1x}}
\end{split}  
\label{eq:B0xSolution}
\end{equation} 
and
\begin{equation}
\begin{split}
B_{0y} &=  \frac{B_{1x} \left(\vec{B}_{0}\cdot\vec{B}_2 - B_{0} B_{2z}\right)- B_{2x}\left(\vec{B}_{0}\cdot\vec{B}_1 - B_{0} B_{1z}\right) }{B_{1x} B_{2y} - B_{2x} B_{1y}}\\
&\approx  \frac{\vec{B}_{0}\cdot\vec{B}_2 - B_{0} B_{2z} }{B_{2x}}
\end{split} ,
\label{eq:B0ySolution}
\end{equation}
where the second lines are obtained by assuming that $B_{1y}$ and $B_{2x}$ are negligible (meaning $\vec{B}_1$ is oriented predominantly along $x$ and $\vec{B}_2$ predominantly along $y$).
It is worth noting here that the statistical uncertainties on $B_{0x}$ and $B_{0y}$ originate from the terms proportional to the scalar products, whereas the accuracy is determined by the terms proportional to $B_{0}B_{1z}$ and $B_{0} B_{2z}$.

\subsubsection{Magnetometric sensitivity }
\label{sec:VarioSensitivity}
%
Based on the second line of Eqs.~\eqref{eq:B0xSolution} and \eqref{eq:B0ySolution}, the statistical uncertainty of the variometer method is determined by
\begin{equation}
\sigma(B_{0\textrm{j}}) = \frac{\sigma(\vec{B}_{0}\cdot\vec{B}_{\textrm{T}}) - \sigma(B_{0}) B_{\textrm{T}z}}{B_{\textrm{T}j}} ,
\label{eq:VarioSens}
\end{equation}
with $j$ indicating the direction of the transverse coil producing $\vec{B}_{\textrm{T}}$.
The components of $\vec{B}_{\textrm{T}}$ do not introduce a statistical uncertainty, as they are fixed by the magnetic field maps.
The precision with which the scalar product between $\vec{B}_{0}$ and $\vec{B}_{\textrm{T}}$ can be determined depends on the amplitude and the duration of the currents applied to the transverse coils.
Let us consider the case of a sequence of $n$ steps of equal duration \hl{$t_s$} with applied currents $I_i$, assuming an anti-symmetric sequence of currents: $\sum_i I_i = 0$.
The uncertainty on the square of the magnetic field magnitude during one step is then given by \hl{$\sigma(B^2) = 2 B \sigma(B) = 2 B \rho(B)/\sqrt{2 t_s}$}, with $\rho(B)$ the noise density of the magnitude (Eq.~\eqref{eq:sens2}).
Using weighted linear least squares fitting, the uncertainty on the coefficient of the linear term in Eq.~\eqref{eq:Parabola} is given by
\begin{equation}
\sigma(2\vec{B}_{0}\cdot\vec{B}_{\textrm{T}}) 
= \frac{\sigma(B^2)}{\sqrt{\sum\limits_{i=1}^n I_i^2}}
= \frac{2 B\, \rho(B)}{ \sqrt{2 t_s}\, \sqrt{\sum\limits_{i=1}^n I_i^2} } .
\label{eq:ScalarProductUncertainty}
\end{equation}
As $ B_{\textrm{T}z}$ is typically not larger than a few nT, the uncertainty on the scalar product $\sigma(\vec{B}_{0}\cdot\vec{B}_{\textrm{T}})$ is about a factor of 1000 larger than $\sigma(B_{0}) B_{\textrm{T}z}$, hence one can neglect the second term in Eq.~\eqref{eq:VarioSens}.
The uncertainty during one measurement cycle is then 
\begin{equation}
\sigma(B_{0\textrm{j}}) 
= \frac{ B}{B_{\textrm{T}j}}\frac{\rho(B)}{ \sqrt{2 t_s}}\frac{1}{\sqrt{\sum\limits_{i=1}^n I_i^2} } .
\label{eq:B0Tuncertainty}
\end{equation}
Taking into account that two transverse projections are needed, the duration of one full variometer measurement cycle is \hl{$2n\,t_s$,} hence giving the following  noise density:
\begin{equation}
\rho(B_{0\textrm{j}}) 
=\sigma(B_{0\textrm{j}})\sqrt{4n\,t_s} = \rho(B)\frac{ B}{B_{\textrm{T}j}}\frac{\sqrt{2n}}{\sqrt{\sum\limits_{i=1}^n I_i^2} } .
\label{eq:B0TnoiseDensity}
\end{equation}
It is clear that in order to get the best sensitivity, one has to use the smallest number of steps $n=3$  ($I$, 0 and $-I$) per transverse field direction at the highest possible current $I$.
A typical variometer measurement cycle for the nEDM experiment then consists of applying a sequence of 3 steps of 6\,s per transverse direction with a maximum applied transverse field of 50\,nT.
Such a measurement typically results in an uncertainty of about 10\,pT, which is about a factor of 3 larger than expected from the calculated noise density.
The reason is that, at this level of precision, the stability of the current source is a limiting factor.
The uncertainty on the squared magnitude of the field should thus be modified to 
\begin{equation}
\sigma(B^2) = \sqrt{\bigg(2B\sigma(B)\bigg)^2 + \left(\sigma(I)\frac{\partial B^2}{\partial I}\right)^2}  ,
\end{equation}
such that the \si{\micro A} precision of the current source can be taken into account.

\subsubsection{Stability and accuracy}
\label{sec:VarioAccuracy}

The accuracy of the variometer method is determined by the accuracy of the field maps of $\vec{B}_{1}$ and $\vec{B}_{2}$ at the positions of the Cs{} sensors. 
These maps typically have an inaccuracy of 1\,nT in all three components.
Particularly the inaccuracy of the $z$-component propagates into a systematic error in $B_{0x}$ and $B_{0y}$ through the terms $ B_{0} B_{1z} /B_{1x}$ and $B_{0} B_{2z}/B_{2x}$ of Eqs.~\eqref{eq:B0xSolution} and \eqref{eq:B0ySolution} respectively.
Using typical values of \SI{1}{\micro T} for $B_{0z}$ and 50\,nT for  $B_{1x}$ and  $B_{2y}$, the estimated accuracy is \hl{20\,nT for $B_{0x}$ and $B_{0y}$.
However, $B_{0z}$ can be determined much more accurately as it is well approximated by the (directly-measured) magnitude $\|\vec{B}_0\|$.}

If the transverse components remain smaller than 10\,nT, as is typically the case in the nEDM experiment, the error made with this approximation is less than 100\,pT.

Luckily, the inaccuracy due to the $B_{0z} B_{Tz}/B_{Tj}$ term is canceled when comparing two variometer measurements of similar main magnetic fields. 
Assuming the main magnetic field direction is not changed too much, the difference between two magnetic fields can be determined with a relative accuracy of a few percent, since the main contribution of $B_{0z}$ to the 20\,nT cancels out when taking a difference.
This of course does not hold when inverting the magnetic field direction. 
As shown in Section \ref{sec:Applications}, these relative measurements are very useful for characterizing drifts of the main magnetic field and provide access to higher order magnetic field gradients that are inaccessible with the regular phase-feedback mode.

\section{Applications of the magnetometer array}
\label{sec:Applications}
The Cs{} magnetometer array can be used for a variety of applications. 
The remainder of this paper will focus on two important ones directly beneficial to the nEDM experiment: (i)~the implementation of a strategy to correct for the drift of the vertical magnetic field gradient and (ii)~a procedure to optimize the homogeneity of the magnetic field.
Section \ref{sec:GzExtraction} describes how to extract the magnetic field gradients from the magnetometer array when vector or scalar magnetic field information is collected.
This procedure is then applied to the data taken during the nEDM experiment to characterize the typical gradient drifts and to estimate the accuracy of the gradient extraction that is solely based on the magnitude readings.
Section \ref{sec:T2opt} outlines the optimization procedure that significantly improved the sensitivity of our nEDM experiment during the 2015 and 2016 data taking campaigns.
%

\subsection{Spatial field distribution and gradient extraction}
\label{sec:GzExtraction}
%
\hl{
In order to extract the relevant magnetic field gradients, we model the spatial field distribution using a multipole expansion.
The multipoles were chosen such that the relevant gradients can be described by a small number of expansion coefficients. 
Specifically, we use the multipole expansion as presented in \cite{Abe19a}, where the magnetic field at position $\vec{r}$  is expanded in the form:
}
\begin{equation}
\vec{B}(\vec{r}) = \sum\limits_{l,m} G_{l,m}
\begin{bmatrix}
\Pi_{x,l,m} (\vec{r})\\
\Pi_{y,l,m}(\vec{r}) \\
\Pi_{z,l,m}(\vec{r})
\end{bmatrix}  ,
\end{equation}
with the $\vec{\Pi}_{l,m}$ harmonic polynomials of degree $l$ in the Cartesian coordinates $x$, $y$ and $z$, and $G_{l,m}$ the corresponding gradient coefficients.
Each degree $l$ has $2l+3$ polynomials, with $m$ ranging from $-(l+1)$ to $l+1$.
The origin of the coordinate system is chosen at the center of the cylindrical precession chamber, as this significantly simplifies averaging over the chamber volume.
The harmonic polynomials up to third order are listed in Table II of \cite{Abe19a}.
%

\hl{The gradient $G_{\textrm{grav}}$ (introduced in Eq.~\eqref{eq:Rratio}), relevant for the nEDM experiment, is a specific combination of the harmonic coefficients \cite{Abe19a}:}
\begin{equation}
G_{\textrm{grav}} = G_{1,0} + G_{3,0} \left(\frac{3H^2}{20}-\frac{3R^2}{4}\right),
\label{eq:GGrav}
\end{equation}
where $H$ is the height and $R$ the radius of the cylindrical storage chamber. 
\hl{Evaluating this expression with} the dimensions of the nEDM precession chamber, the vertical gradient is given by $G_{\textrm{grav}} = G_{1,0} - G_{3,0} (393\textrm{\,cm}^2)$.

\subsubsection{Gradient extraction in the variometer mode}
\label{sec:GzVario}
If the vector components of the magnetic field are known at positions $\vec{r}_i$, the gradients $G_{l,m}$ can be determined by solving the matrix equation  
\begin{equation}
\begin{bmatrix}
{B_x}\\
{B_y}\\
{B_z} 
\end{bmatrix}
= 
\begin{bmatrix}
{\Pi_x}\\
{\Pi_y}\\
{\Pi_z}\\
\end{bmatrix}
{G} ,
\label{eq:GradientMatrixEq}
\end{equation}
where ${B_x}$ is a column vector with elements $B_{x}^i$ representing the $x$-component of the magnetic field measured at positions $\vec{r}_i$, ${\Pi_x}$ is a matrix with elements $({\Pi_x})^{ij} = \Pi_{x,l_j,m_j}(\vec{r}_i)$, i.e., the harmonic polynomial defined by $l_j$ and $m_j$ evaluated at position $\vec{r}_i$, and ${G}$ is a column vector containing the harmonic coefficients $G_{l_j,m_j}$.
The expressions are similar for the $y$- and $z$-matrices.
In the particular case of measurements with the variometer method there is, however, a significant difference between the uncertainty on $B_z$ and the uncertainties on the transverse components $B_x$ and $B_y$. Therefore, each line in the matrix equation is weighted with the inverse of the squared uncertainty of the corresponding magnetic field component value.
%


Since one of the HV-compatible magnetometers failed after an electrical discharge burned one of its optical fibers at an early stage of data taking, we only have 15 sensors available to fit the harmonic coefficients.
This results in $3\times15 = 45$ equations, enabling us to comfortably fit up to third order (24 harmonics) while still having enough degrees of freedom for error estimation.
This means that the harmonic coefficients necessary for the estimation of $G_{\textrm{grav}}$ are easily accessible using the variometer method.
However, since the method involves applying additional magnetic fields, it is not used during a typical nEDM measurement cycle as it would disturb the neutron EDM measurement.
%

\subsubsection{Gradient extraction in the phase-feedback mode}
\label{sec:GzPLL}
%

Since in phase-feedback mode only the magnitude of the magnetic field is known at positions $\vec{r}_i$, we first have to make the following approximation:
\begin{equation}
\pm\|\vec{B}\|=B_z+ \frac{B^2_x+B^2_y}{2B_z}+\cdots\approx B_z ,
\end{equation}
where the sign is determined by the main direction of $\vec{B}$, which is oriented along the $z$-axis. 
This approximation is valid in the nEDM experiment as the field maps have shown that the transverse components of the main \SI{1}{\micro T} field are typically smaller than \SI{10}{nT}. 
To extract the magnetic field gradients $G_{l,m}$, one has to solve the matrix equation ${B_z} = {\Pi_z^{\textrm{S}}}{G^{\textrm{S}}}$, with the matrices being defined as in Eq.~\eqref{eq:GradientMatrixEq}, with the exception that the polynomials with $m_j=\pm(l_j+1)$ are not included. 
The reason for this is that these modes are purely transverse and do not contribute to $B_z$, and are therefore not accessible via the magnitude.
The superscript S (scalar) is added to make a clear distinction between gradients ${G}$ determined from vector measurements and  gradients ${G^{\textrm{S}}}$ extracted from scalar measurements.
Again, the uncertainty on the magnitude measurements can be used to assign weights to the equations.
%

As the nEDM experiment has sixteen Cs{} magnetometers, we typically limit the scalar harmonic expansion to second order (with 9 fit parameters), providing the following magnetic field description:
\begin{align}
B_z(x, y, z) \, =\, & G_{0,0}^{\textrm{S}} + y\, G_{1,-1}^{\textrm{S}}+ z\,G_{1,0}^{\textrm{S}} +x\, G_{1,1}^{\textrm{S}} 
\nonumber
\\
 & + 2xy\,G_{2,-2}^{\textrm{S}}+2yz\,G_{2,-1}^{\textrm{S}}
  \nonumber
 \\
&+\left(z^2-\frac{1}{2}(x^2+y^2)\right)\, G_{2,0}^{\textrm{S}} 
 \nonumber
 \\
&+ 2xz\,G_{2,1}^{\textrm{S}} + (x^2-y^2)\,G_{2,2}^{\textrm{S}}  ~ .
\label{eq:polynomial}
\end{align}
\hl{The cubic vertical gradient $G_{3,0}$ clearly cannot be determined using Eq.~(\ref{eq:polynomial}). }
However, the higher order terms do affect the extracted scalar gradients ${G^{\textrm{S}}}$.
Assuming a multipole expansion ${B_z}={\Pi_z} {G}$, the contribution of the higher order terms to the scalar fit parameters can be calculated explicitly:
\begin{equation}
{G^{\textrm{S}}} = \left(\left({\Pi_z^{\textrm{S}}}\right)^{\textrm{T}} W \, {\Pi_z^{\textrm{S}}}\right)^{-1} \left({\Pi_z^{\textrm{S}}}\right)^{\textrm{T}} W \,{\Pi_z} \,{G} ,
\end{equation}
where $W$ is a diagonal matrix containing the weight of each equation.
Using the positions of the 15  Cs{} magnetometers that were operational during the 2015/2016 nEDM data taking and assuming equal weights for each magnetometer, the influence of the third order gradients on the vertical linear gradient $G_{1,0}^{\textrm{S}} = \sum a_{l,m} G_{l,m}$ is summarized in Table \ref{tab:G10Scubic}.
%
%
\hl{By comparing the prefactors in the definition of $G_{\textrm{grav}} = G_{1,0} - \SI{393}{cm^2} \,G_{3,0}$ in Eq.~\eqref{eq:GGrav}  to the prefactors $a_{1,0}{=}1$ and $a_{3,0}{=}\SI{-288}{cm^2}$, we can conclude that $G^\textrm{S}_{1,0}$ is a reasonable but slightly inaccurate estimator for $G_{\textrm{grav}}$.
Adding weights $W$ based on the typical uncertainties of each sensor changes the factors $a_{3,m}$ in Table \ref{tab:G10Scubic}, but the prefactor for $G_{3,0}$ remains about 3/4 of the factor in $G_{\textrm{grav}}$. 
}

\subsubsection{Gradient extraction during nEDM data taking}
\label{sec:TypicalGz}
%
In order to show that the Cs{} magnetometer array meets the requirements for gradient drift correction outlined at the end of Section \ref{sec:nEDMexperiment}, we have to quantify the sensitivity and accuracy of the gradient extraction procedure based on the real magnetic field conditions in the nEDM experiment.
To monitor the magnetic field during nEDM data taking, the typical measurement procedure regarding the Cs{} sensors consists of: (i)~calibrations before and after each nEDM run to monitor the light intensity and the reference phase of the phase-feedback mode, (ii)~followed by variometer measurements to monitor the higher order gradient drifts, and (iii)~continuous measurements in the phase-feedback mode during the nEDM run.
 A run typically takes a few days, corresponding to about 500 Ramsey cycles which each take five minutes, while the electric field is reversed every 56 cycles.
 %

\begin{figure}
\centering
\includegraphics[width=\linewidth]{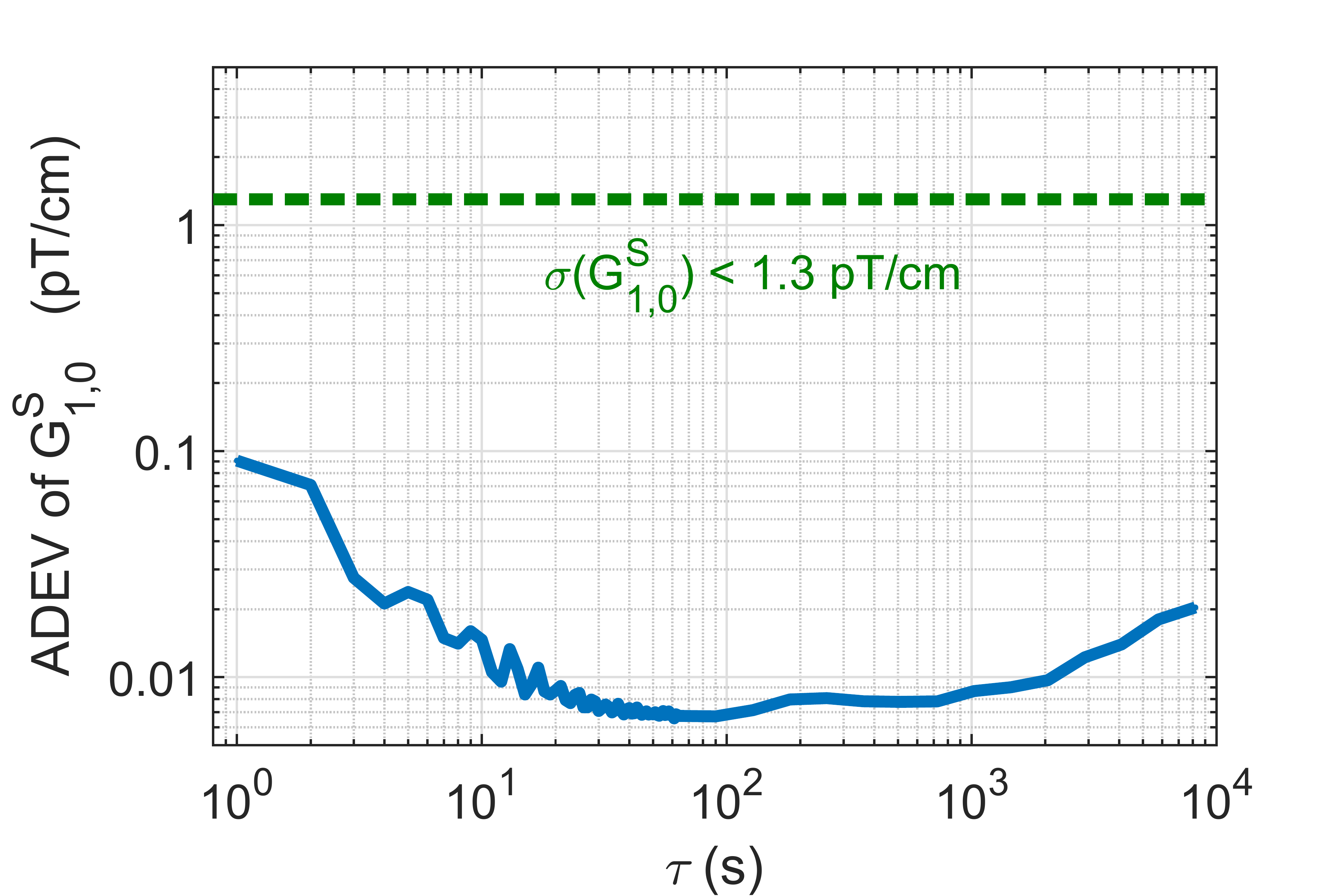}
\caption{The Allan deviation of the vertical gradient $G_{1,0}^{\textrm{S}}$ extracted from the data shown in Fig.~\ref{fig:ASDBz} using the model in Eq.~\eqref{eq:polynomial} is shown in blue. The achievable statistical uncertainty at the nEDM cycle duration of $180$\,s is 8\,fT/cm, which is significantly below the upper limit indicated as a dashed green line. Statistical uncertainties in the magnetometers cause the rising slope towards small $\tau$ values. The result at $180$\,s is not limited by the slope but rather by the stability of the measurement system.}
\label{fig:ASDGz}
\end{figure}

In order to quantify the gradient drift sensitivity during a Ramsey cycle, we extract $G_{1,0}^{\textrm{S}}$ from the data used in Fig.~\ref{fig:ASDBz} and calculate its Allan  deviation (ADEV).
The results are shown in Fig.~\ref{fig:ASDGz}.
It is clear that the realized gradient sensitivity during the neutron storage time of 180\,s is significantly better than the requirement of 1.3\,pT/cm calculated in Section \ref{sec:nEDMexperiment}. 
The ADEV slowly increases for longer integration times but remains far below the limit for all relevant time-scales. 

Regarding the accuracy of the gradient drift measurement, there are two effects that play a role. 
On the one hand there are sensor-related drifts that translate into an artificial gradient drift, on the other hand there are drifts of $G_{1,0}^{\textrm{S}}$ induced by changes in the higher order magnetic field gradients.
To estimate the former, we compare the calibrations before and after each nEDM run, to characterize the latter, we use the variometer measurements.

As discussed in Section \ref{sec:PLLaccuracy}, the typical change in reference phase between calibration pairs before and after an nEDM run results in reading offsets corresponding to a few pT.
Using the magnetic field gradient model of Section \ref{sec:GzPLL}, these offsets produce a change of the fit-parameter $G_{1,0}^{\textrm{S}}$ with a standard deviation of 0.1\,pT/cm in the time span of a few days.
Similarly, the light intensity changes slightly over the course of an nEDM run, modifying the light shift in each sensor, resulting in an artificial gradient drift with standard deviation of 0.03\,pT/cm.
Since the light intensity typically decreases over time and the direction of the laser beams is opposite for the sensors above and below the storage chamber, the average change is about -0.01\,pT/cm.

Comparing the variometer measurements before and after each nEDM run, we can extract the total change of each gradient $\Delta G_{l,m}$ during the run.
The distribution of $\Delta G_{l,m}$ is Gaussian, with the standard deviation of the terms relevant to $G_{1,0}^{\textrm{S}}$ summarized in Table~\ref{tab:G10Scubic}.
Taking into account the correlation between the drifts of $G_{1,0}$ and $G_{3,0}$, and using Eq.~\eqref{eq:GGrav}, the standard deviation of $\Delta G_{\textrm{grav}}$ is 1.4\,pT/cm.
Using the magnitude of the same data, the extracted drift of $G_{1,0}^{\textrm{S}}$ is in agreement with the drift of $G_{\textrm{grav}}$ within the error bars of the parameters, which are typically 0.7\,pT/cm for $G_{\textrm{grav}}$ due to the inaccuracy of the variometer mode \hl{including map-related inaccuracies}.
This gives an upper limit on the relative accuracy of $G_{1,0}^{\textrm{S}}$: the accuracy is at least a factor of 2 better than the standard deviation of the drift on the timescale of an nEDM run.
It follows that the dominant uncertainty on the extracted gradients is not due to the accuracy of the individual sensors, but rather due to the `aliasing effect' of the higher order modes which are not included in the fit.
\begin{table}
\setlength{\tabcolsep}{8pt}
\caption{Overview of the harmonic coefficients contributing to the fit parameter $G_{1,0}^{\textrm{S}} = \sum a_{l,m}G_{l,m}$ up to degree 3. 
For each harmonic coefficient $G_{l,m}$, the weighing factor $a_{l,m}$ and the standard deviation of the gradient drift $\sigma(\Delta G_{l,m})$ during a typical nEDM run are given. 
Taking into account the correlations between the different contributions, an estimation of the standard deviations of the drift of $G_{1,0}^{\textrm{S}}$ and its accuracy $G_{1,0}^{\textrm{S}} - G_{\textrm{grav}}$ are given in the last two lines.
\hl{In the last line, the error estimation is scaled with $\sqrt{\chi^2/\nu}$ of the variometer fit to take into account the map-related inaccuracies of the method.}
}
\centerline{
\begin{tabular}{l c c }
\hline
\hline
$G_{l,m}$& $a_{l,m}$ (cm$^{l-1})$ & $\sigma(\Delta G_{l,m})$ (pT/cm$^l)$ \\
\hline
$G_{1,0}$ &1&1.71 \\
$G_{3,-3}$ &-135&0.0009 \\
$G_{3,-2}$ &344& 0.0006\\
$G_{3,-1}$ &22& 0.0015\\
$G_{3,0}$ &-288& 0.0023\\
$G_{3,1}$ &-23& 0.0010\\
$G_{3,2}$ &466&0.0010 \\
$G_{3,3}$ &1& 0.0017\\
\hline
$G_{1,0}^{\textrm{S}}$&&1.4--1.7\\
$G_{1,0}^{\textrm{S}} - G_{\textrm{grav}}$ &&$<0.7$\\
 \hline
 \hline
 \end{tabular}
}
\label{tab:G10Scubic}
\end{table}

\subsection{Homogenization of the magnetic field}
\label{sec:T2opt}
%
The homogeneity of the magnetic field influences both the statistical precision of the nEDM experiment and its systematic effects.
To improve the former without exacerbating the latter, we have developed a procedure for optimizing the magnetic field in the precession chamber.
%
%
The principles behind this optimization strategy are explained in Section \ref{sec:T2optStrategy}.
The implementation of the routine is described in Section \ref{sec:T2optImpl}, followed by a discussion of the tuning of the algorithm in Section \ref{sec:T2optTuning}.
Finally, the resulting improvement in sensitivity is presented in Section \ref{sec:T2optResults}.
%

\subsubsection{Principles behind the optimization}
\label{sec:T2optStrategy}
%
Improving the statistical sensitivity and minimizing the systematic effects impose different requirements on the magnetic field optimization.
%
The magnetic-field-related contribution to the statistical precision of the nEDM measurement is captured in the parameter $\alpha$ of Eq.~\eqref{eq:nustatistical}, which is the visibility or contrast of the Ramsey resonance.
This parameter is predominantly defined by the neutrons' transverse spin relaxation time $T_2$ via $\alpha(T) = \alpha_0 \exp{(-T/T_2)}$ where $\alpha_0$ is the polarization at the start of the Ramsey procedure and $T$ the precession time of the neutrons.
The transverse relaxation time results from a combination of three types of neutron depolarization in the storage chamber, as discussed in \cite{Abe19a}.
The first mechanism is depolarization due to wall collisions, which is an effect that does not depend on the magnetic field.
The second is gravitationally enhanced depolarization \cite{GravEnhDepol1,GravEnhDepol2}, which is caused by the extremely low kinetic energy of the ultracold neutrons.
Different energy groups of neutrons have a different average height in the chamber, \hl{so in the presence of a vertical gradient of the field's main component their precession frequencies differ slightly.}
This causes a dephasing of the different energy groups, which results in a lower polarization at the end of the Ramsey procedure.
To reduce this effect, \hl{it is crucial to minimize specifically the vertical gradient  $\partial B_{z}/\partial z$}.
The third mechanism is intrinsic depolarization, which refers to the depolarization within each given energy group.
Even though the neutrons have the same energy, their trajectories through the chamber differ, resulting in dephasing if the magnetic field is not homogeneous over the chamber volume.
\hl{Such local changes in Larmor frequency are caused by all gradients of the main field component $B_{z}$ while gradients of the transverse components $B_{x}$ and $B_{y}$ play a negligible role.}

Conversely, the magnetic-field-related systematic effects that are not dealt with in the extension of the {\it crossing point analysis} of \cite{Abe19a}, involve the quantities $\langle B_{\textrm{T}}^2\rangle$  and $G_{3,0}$.
The first is defined as
\begin{equation}
\langle B_{\textrm{T}}^2\rangle = \Big\langle \big(B_x-\langle B_x\rangle\big)^2+\big(B_y-\langle B_y\rangle\big)^2\Big\rangle\
\end{equation}
and stands for the square of the transverse magnetic field components averaged over the storage volume. 
It is a second order combination of the harmonic expansion coefficients $G_{l,m}$:
\begin{equation}
\langle B_{\textrm{T}}^2\rangle = \sum a_{ij} G_{l_i,m_i}G_{l_j,m_j} .
\end{equation}
The coefficients $a_{ij}$ are given in Appendix B of \cite{Abe19a}.
The smaller the gradients of the transverse magnetic field components, the smaller this systematic effect.

\hl{The quantity $G_{3,0}$ is the cubic vertical gradient of $B_z$ with a characteristic $z$-dependance $\vec{B}(x{=}0,y{=}0,z)\propto (0,0,z^3)$.
The systematic uncertainties related to $G_{3,0}$ can thus be suppressed by ensuring the homogeneity of $B_z$}

\hl{In summary, optimizing the homogeneity of the longitudinal field component $B_z$ helps to suppress certain systematic uncertainties and is crucial to maintain long $T_2$ times and thus a high statistical sensitivity. 
Optimizing the homogeneity of the transverse field components $B_x$ and $B_y$ is equally important since a different systematic effect is related to those components. 
}

\subsubsection{Implementation}
\label{sec:T2optImpl}
%
\hl{Firstly, the homogeneity of the longitudinal magnetic field component $B_z$ can be directly accessed by the Cs{} magnetometer array}.
However, since the sensors are not perfectly accurate and require offline corrections, $B_z$ was only available up to an accuracy of about 45 to 90\,pT during online data taking (Table~\ref{tab:Systematics}).
Therefore, the goal of the optimization routine is to reduce the spread of the Cs{} magnetometer readings to this level.
Secondly, the transverse components are accessible with the variometer method, but the accuracy is not sufficient to keep $\langle B_{\textrm{T}}^2\rangle$ below the goal of 2\,nT$^2$, which would correspond to a systematic effect at the level of a few 10$^{-27}\ecm$.
\hl{For this reason, offline field maps, that were recorded before the period of nEDM data taking,  are used to provide an estimate of $\langle B_{\textrm{T}}^2\rangle$.}
The final correction of this systematic effect will be performed with more accurate values extracted from a more recent mapping campaign (the analysis of which will be included in the third part of the trilogy).

Combining the online information of the Cs{} sensors with the offline magnetic field maps, we developed a routine to optimize the currents $I_{\textrm{coil}}$ applied to a set of 30 trim-coils wound around the vacuum tank. 
The magnetic field produced by each coil when applying one unit of current was characterized both online and offline, providing $\vec{B_{\textrm{coil}}^{\textrm{CsM}}}$ measured by the Cs{} magnetometer (CsM) in the variometer mode, and the harmonic expansion coefficients $G_{\textrm{coil}}^{\textrm{map}}$ as extracted from the magnetic field maps.
After measuring the main magnetic field $\vec{B_{0}^{\textrm{CsM}}}$  on-line with the Cs{} magnetometer array, the optimal currents are calculated by minimizing the sum of the following three terms:
\begin{equation}
\label{eq:SumToMin}
S =  S_{\textrm{Long}}+ T_{\textrm{Trans}}  S_{\textrm{Trans}} +T_{\textrm{Reg}} S_{\textrm{Reg}} ,
\end{equation}
where  $S_{\textrm{Long}}(I_{\textrm{coil}})$ quantifies the homogeneity of the longitudinal component,  $S_{\textrm{Trans}}(I_{\textrm{coil}})$ evaluates the systematic effect due to the transverse components and $S_{\textrm{Reg}}(I_{\textrm{coil}})$ is added as a regularization term since there are more parameters than constraints (30 $>$ 16+1).
The factors $T_{\textrm{Trans}}$ and $T_{\textrm{Reg}}$ are tuning parameters and assign a weight to the respective sums relative to $S_{\textrm{Long}}$.
%

The explicit expression for $S_{\textrm{Long}}$ as a function of the currents $I_{\textrm{coil}}$ is given by
\begin{equation}
S_{\textrm{Long}} = \sum\limits_{\textrm{CsM}}\left(B_{0,z}^{\textrm{CsM}}  + \sum\limits_{\textrm{coil}}I_{\textrm{coil}} B_{\textrm{coil},z}^{\textrm{CsM}} - B_{\textrm{goal}}\right)^2 ,
\end{equation}
where $B_{0,z}^{\textrm{CsM}}$ and $B_{\textrm{coil},z}^{\textrm{CsM}}$ are the $z$-components measured by the Cs{} magnetometer of the main magnetic field and the field produced by the coil when applying one unit of current respectively. $B_{\textrm{goal}}$ is the goal value for the Cs{} sensor magnitude readings.
Typically, the sensors are all assigned the same goal value to improve the homogeneity, but other configurations are possible.
%

The transverse requirements are taken into account by the following sum
\begin{equation}
S_{\textrm{Trans}} = \langle (B_{\textrm{T}}^{\textrm{map}})^2 \rangle = \sum\limits_{i,j} a_{ij} G^{\textrm{map}}_{l_i,m_i}G^{\textrm{map}}_{l_j,m_j} ,
\end{equation}
where $G_{l_i,m_i}^{\textrm{map}} = G_{0,l_i,m_i}^{\textrm{map}} + \sum\limits_{\textrm{coil}} I_{\textrm{coil}} G_{\textrm{coil},l_i,m_i}^{\textrm{map}}$ is the harmonic coefficient $G_{l_i,m_i}$ of the total magnetic field that would be produced if the currents $I_{\textrm{coil}}$ would be applied to the coils as determined from the field maps. 
The coefficients $a_{ij}$ are defined in \cite{Abe19a}.
%

The regularization term is given by
\begin{equation}
S_{\textrm{Reg}} = \sum\limits_{\textrm{coil}} \left(I_{\textrm{coil}}\max_{\textrm{CsM}}(\| \vec{B^{\textrm{CsM}}_{\textrm{coil}}}\|)\right)^2 ,
\end{equation}
where $\max\limits_{\textrm{CsM}}(\| \vec{B^{\textrm{CsM}}_{\textrm{coil}}}\|)$ is the maximum magnitude measured by the Cs{} magnetometers when one unit of current is applied to the coil.
This term makes sure that the \hl{magnetic field produced} per coil is not too large, avoiding a loss in sensitivity due to local inhomogeneities created by the coils themselves.

In order to minimize Eq.~\eqref{eq:SumToMin}, we solve the set of equations $\partial S/\partial I_{\textrm{coil}} = 0$.
Since the terms in $S$ are at most of order 2 in $I_{\textrm{coil}}$, $\partial S/\partial I_{\textrm{coil}}$ is of order 1 and can be solved efficiently using matrix inversion.

\subsubsection{Optimizing the tuning parameters}
\label{sec:T2optTuning}
%
The success of the algorithm is determined by the choice of the tuning parameters $T_{\textrm{Trans}}$ and $T_{\textrm{Reg}}$.
To determine the optimal values, we start off with an estimate of the optimal size of each sum in Eq.~\eqref{eq:SumToMin}.
Given the on-line accuracy of the Cs{} magnetometers, we estimate the final standard deviation of $(B_{z}^{\textrm{CsM}}-B_{\textrm{goal}})$ at 100\,pT, resulting in a longitudinal term $S_{\textrm{Long}}$ of $(0.1\,\textrm{nT})^2\times16 = 0.16$\,nT$^2$.
The value of $\langle B_{\textrm{T}}^2\rangle$ should be as small as possible, but since the maps provide only a rough estimate, we set the goal value for $S_{\textrm{Trans}}$ at 0.5\,nT$^2$.
To avoid producing local inhomogeneities due to strong currents in the trim-coils, the tuning is started with a trial value of 2\,nT produced per coil on average, resulting in a regularization term $S_{\textrm{Reg}} $ of $ (2\,\textrm{nT})^2\times 30 = 120$\,nT$^2$.
Comparing the size of each sum, first guesses for the tuning parameters are $T_{\textrm{Trans}} = S_{\textrm{Long}}/S_{\textrm{Trans}} = 0.32$ and $T_{\textrm{Reg}} = S_{\textrm{Long}}/S_{\textrm{Reg}} = 0.0013$.

\begin{figure}
\centering
\includegraphics[width=0.8\linewidth]{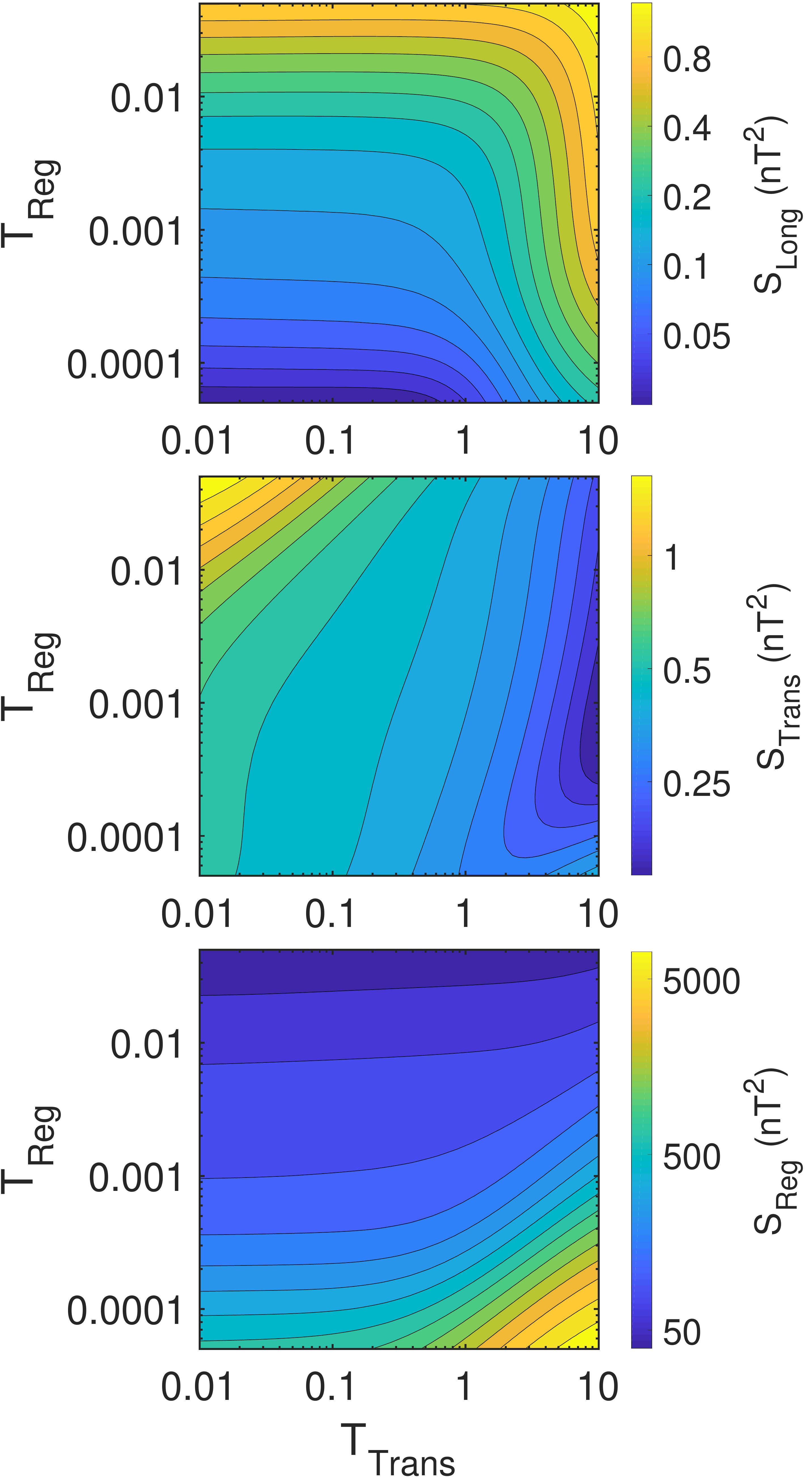}
\caption{The behavior of $S_{\textrm{Long}}$ (top),  $S_{\textrm{Trans}}$ (middle) and $S_{\textrm{Reg}}$ (bottom) evaluated at the optimal trim-coil currents as a function of the tuning parameters $T_{\textrm{Trans}}$ and $T_{\textrm{Reg}}$. All scales (including the color scale) are logarithmic.}
\label{fig:TuningParameters}
\end{figure}

Figure~\ref{fig:TuningParameters} shows the minimized values of each sum $S_i$ in Eq.~\eqref{eq:SumToMin} in function of the tuning parameters, with the ranges centered around our initial guesses.
The terms are calculated using a typical magnetic field which is measured on-line 30 minutes after degaussing the $\mu$-metal shield, as is the typical procedure during nEDM data taking.
As is clearly visible in the two uppermost plots of Fig.~\ref{fig:TuningParameters}, the tuning parameter $T_{\textrm{Trans}}$ (horizontal axis) determines the relative importance of the longitudinal spread (top) versus the transverse homogeneity (middle).
For values of $T_{\textrm{Trans}}$ smaller than $1.0$, the longitudinal spread is almost solely determined by the regularization parameter $T_{\textrm{Reg}}$.
The smaller $T_{\textrm{Reg}}$, the larger the applied currents (bottom), and the smaller the predicted spread of $B_{z}$.
For $T_{\textrm{Trans}}$ larger than 1, the value of $\langle (B_{\textrm{T}}^{\textrm{map}})^2 \rangle$ is significantly reduced at the cost of a worse \hl{$B_z$ homogeneity} and much larger currents.
The behavior at large $T_{\textrm{Trans}}$ and small $T_{\textrm{Reg}}$ (bottom right corner of each plot) suggests that it is nearly impossible to have both a small spread in on-line $B_{z}^{\textrm{CsM}}$-component and a small $\langle B_{\textrm{T}}^2\rangle$ predicted from the maps, even if the restriction on the applied currents is relaxed.
This indicates that the estimation of $\langle B_{\textrm{T}}^2\rangle$ from the maps is only reliable down to the 0.3\,nT$^2$ level.
%
%
As the exact size of $S_{\textrm{Trans}}$ is not crucial, $T_{\textrm{Trans}}$ is typically fixed at a value smaller than $1.0$ leading to $\langle (B_{\textrm{T}}^{\textrm{map}})^2 \rangle$ values smaller than the limit of 2\,nT$^2$.

The optimal choice for $T_{\textrm{Reg}}$ is not so straightforward.
It depends on the initial homogeneity of the magnetic field, as a larger inhomogeneity implies a larger amount of current necessary to compensate.
Moreover, as the applied currents become larger, the uncertainty on the measurement of  $B^{\textrm{CsM}}_{\textrm{coil},z}$ will make the estimation of the longitudinal spread inaccurate and thus reduce the predictive power for the value of $\alpha$.
On top of that, making the magnetic field magnitude the same at all sensor positions does not mean that the field in the storage chamber itself is homogeneous, especially when the applied trim-coil currents are large.
For this reason, we typically selected a scan range of 0.0002 to 0.0020 for $T_{\textrm{Reg}}$ and picked out the best setting by measuring the resulting $\alpha$ on-line. 
%

\subsubsection{Results}
\label{sec:T2optResults}
\begin{figure}
\centering
\includegraphics[width=\linewidth]{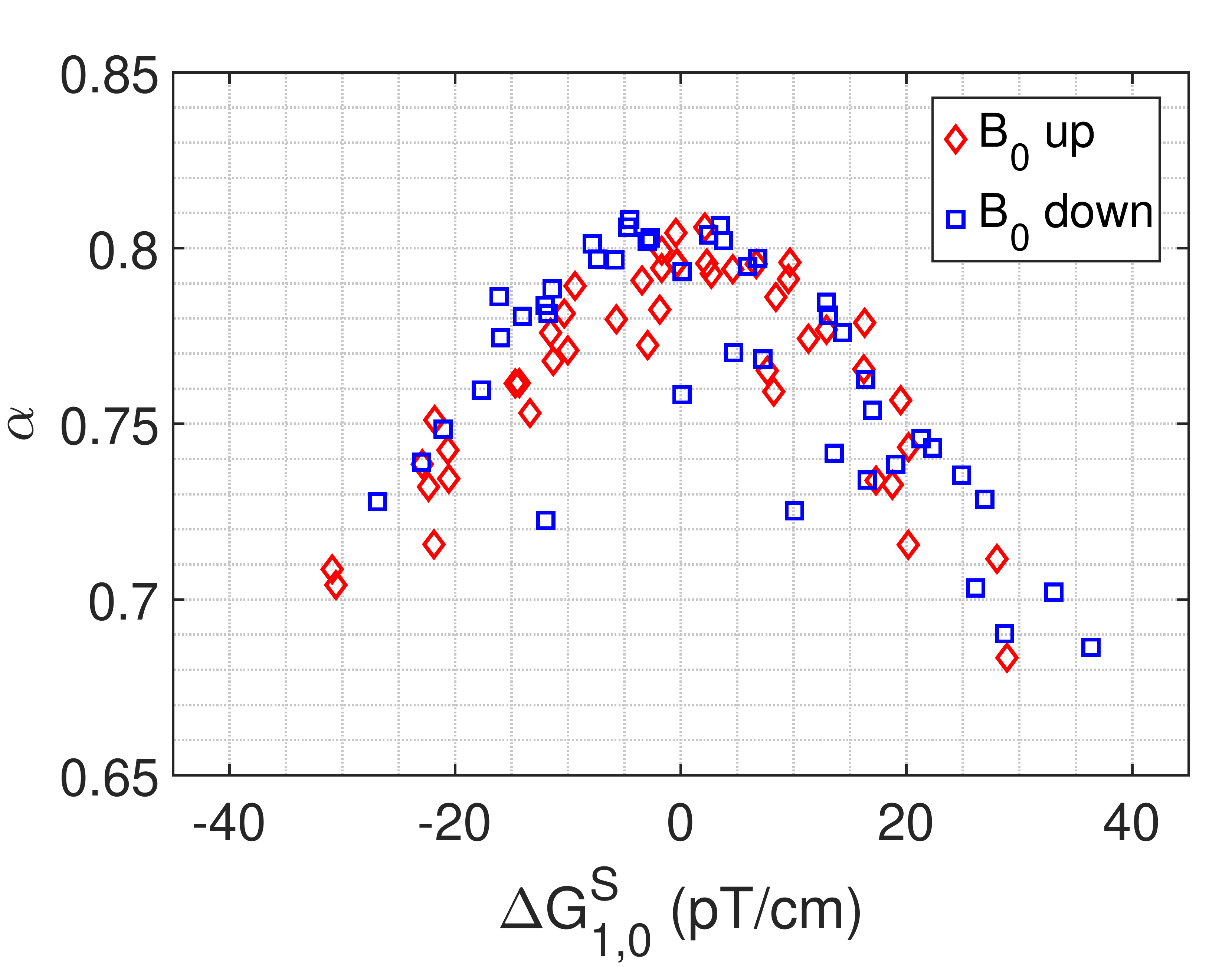}
\caption{The Ramsey contrast or visibility $\alpha$ measured during the nEDM data taking period of 2015 and 2016 as a function of the vertical gradient. The `zero' gradient is defined per magnetic field base configuration (or equivalently per set of measurements that are based on the same homogenization result) as the gradient at which the visibility-parabola reaches its highest point. For nonzero vertical gradient, gravitational depolarization reduces the contrast of the Ramsey curve. Note that both $B_0$ up and $B_0$ down reach similar visibilities.}
\label{fig:alpha}
\end{figure}

%
Different iterations of the optimization procedure were used during the nEDM data taking period of 2015 and 2016. 
For each chosen current setting during data taking, the value of $\langle B_{\textrm{T}}^2 \rangle$ was smaller than 2\,nT$^2$.
The corresponding Ramsey visibilities are shown in Fig.~\ref{fig:alpha}.
The effect of gravitational depolarization is clearly visible as $\alpha$ decreases when the vertical gradient $\Delta G_{1,0}^{\textrm{S}}$ moves away from zero.
From dedicated measurements at different storage times, we know that the initial polarization $\alpha_0$ in our storage bottle is 0.86.
The $\alpha$ values of 0.76-0.81 at zero gradient then correspond to transverse neutron relaxation times between 1450\,s and 3000\,s.

The improvement of the neutron spin relaxation time $T_2$ and the corresponding increase of Ramsey contrast $\alpha$ is summarized in Table~\ref{tab:T2}, comparing data from 2014 without CsM based homogenization with data from 2015 and 2016.
The transverse relaxation time has more than doubled with the new homogenization procedure, resulting in an increase of $\alpha$ by about 35\% and an equal improvement of the nEDM sensitivity.
In order to realize the same improvement with neutron statistics, the total number of detected neutrons would have to be increased by a factor of 1.8 due to the $\sqrt{N}$ scaling (see Eq.~\eqref{eq:nustatistical}).
This is a significant improvement for an experiment that is scheduled to take data for several years.

\begin{table}
\caption{\label{tab:T2} Comparison of the transverse neutron spin relaxation time $T_2$ and the Ramsey contrast $\alpha$ at zero vertical gradient before and after the field homogenization was introduced in 2015. The polarization $\alpha_0$ at the start of the Ramsey procedure is 0.86 in both datasets. In 2014 the $\alpha$ values were significantly different for the two $B_0$ field orientations.}
\centerline{
\begin{tabular}{c c c c}
\hline
\hline
 Year &$B_0$ direction & $T_2$ (s)& $\alpha$ \\
  \hline			
  2014 & up & 760 & 0.64 \\
   &  down & 439 & 0.52 \\
\hline
2015 \& 2016 &  up & 1620-3000 & 0.77-0.81 \\
 &  down & 1450-3000 & 0.76-0.81  \\
  \hline			
  \hline    
\end{tabular}
}
\end{table}

\section{Summary}
We have discussed the design, implementation and performance of the Cs{} magnetometer array installed at the PSI-nEDM experiment.
The compact optical magnetometers are vacuum and HV compatible and are placed on the electrodes above and below the UCN storage chamber, providing on-line gradient information.
The sensors are driven by a single diode laser, using beam multiplexing to bring the light to the individual sensors in the vacuum chamber of the experiment.
We have explained the phase-feedback mode of sensor operation in the $M_x$ configuration and demonstrated an intrinsic magnetometer sensitivity which is below 50\,fT/$\sqrt{\mathrm{Hz}}$ in the shot noise limit.
The final magnetometer noise in the nEDM experiment was significantly larger than the shot noise limit but it did not limit the extraction of important field parameters at the relevant integration time of 180\,s.
At such large integration times the performance is rather limited by system stability which we could demonstrate to be significantly better than required (see Fig.~\ref{fig:ASDGz}). 
We have discussed various systematic effects that influence the reading of the magnetometer and estimated an on-line accuracy of 45 to 90\,pT.
Using a set of two transverse coils, we can run the magnetometers in variometer mode, providing vector information of the local magnetic field.
%

A model was presented to describe the spatial field distribution, and the precision and accuracy of gradient extraction during nEDM data taking was discussed.
Further, a magnetic field homogenization procedure, which more than doubled the transverse spin relaxation time of the neutrons, while at the same time keeping magnetic-field-related systematic effects under control, was presented.
This resulted in an improvement of 35\% of the statistical sensitivity of the nEDM experiment which reduced the time to reach a given statistical sensitivity by a factor of 1.8.

\hl{
The presented techniques are useful in general for the measurement and control of magnetic field uniformity.
We will use an upgraded version of the magnetometer array, based on all-optical sensors \cite{Gru15}, in our new neutron EDM experiment (n2EDM). 
The new sensors use free spin precession in contrast to the driven spin precession in a $M_x$ magnetometer.
This leads to improved stability and accuracy, necessary to fulfill the requirements of our next-generation experiment.
}

\section{Acknowledgements}

We would like to thank the mechanical workshop at the University of Fribourg for manufacturing the construction elements of Cs{} sensors, C.~Macchione for the preparation of the paraffin-coated Cs{} cells and M.~Meier and F.~Burri for their support during the installation of the Cs{}  magnetometer array and measurements at PSI.

The LPC Caen and the LPSC acknowledge the support of the French Agence Nationale de la Recherche (ANR) under Reference No. ANR-09-BLAN-0046. 
P.M. likes to acknowledge support from SNSF-FCS Grant No. 2015.0594 (ETHZ).
E.W. acknowledges the fellowship of the Fund for Scientific Research Flanders (FWO).
This research was partly financed by the Fund for Scientific Research, Flanders;
Grant No. GOA/2010/10 of KU Leuven; the Swiss National Science Foundation projects 126562 (PSI), 140421 (UNIFR), 144473 (PSI), 149211 (ETH), 162574 (ETH), 172626 (PSI), 172639 (ETH), and 181996 (Bern), 
the Deutsche Forschungsgemeinschaft projects BI 1424/2-1 and /3-1, and Grants No. ST/K001329/1, No. ST/M003426/1, and No. ST/L006472/1 from the Science and Technology Facilities Council (STFC) of the United Kingdom. The original nEDM apparatus without the Cs magnetometer array was funded by grants from the PPARC (now STFC) of the United Kingdom. Our Polish partners wish to acknowledge support from the National Science Centre, Poland, under Grants No. UMO-2015/18/M/ST2/00056 and UMO-2016/23/D/ST2/00715.


%

\end{document}